\documentclass[twoside,floatfix,12pt]{mbook}
\usepackage{graphicx}
\usepackage{amssymb}
\usepackage{amsmath}
\usepackage[T1]{fontenc}
\usepackage[cp 1250]{inputenc}
\usepackage{textcomp}
\usepackage{amsfonts}
\usepackage{epsfig}
\usepackage{here}
\usepackage{cite}
\usepackage{esint}
\usepackage{fancyhdr}
\usepackage{color}
\title{{\bf Analysis of the differential cross sections for the reaction $pp\rightarrow ppK^{+}K^{-}$ in view of the $K^{+}K^{-}$ interaction }}
\author{Michał Silarski}
\date{11 sierpnia 2006 r.}
\begin{document}
\thispagestyle{empty}
\newpage
\thispagestyle{empty}
\begin{center}
{\large INSTITUTE OF PHYSICS  }\\
{\large FACULTY OF PHYSICS, ASTRONOMY  }\\
{\large AND APPLIED COMPUTER SCIENCE}\\
{\large JAGIELLONIAN UNIVERSITY}\\
\end{center}
\begin{center}\end{center}
\begin{center}\end{center}
\begin{center}
{\Large{\bf Analysis of the differential cross sections}}
\end{center}
\begin{center}
{\Large{\bf for the reaction $pp\rightarrow ppK^{+}K^{-}$}}
\end{center}
\begin{center}
{\Large{\bf in view of the $K^{+}K^{-}$ interaction}}
\end{center}
\begin{center}\end{center}
\begin{center}
{\large Michał Silarski}
\end{center}
\begin{center}\end{center}
\begin{center}\end{center}
\begin{center}
{\normalsize Master Thesis  }\\
{\normalsize prepared in the Nuclear Physics Division\\}
{\normalsize of the Jagiellonian University\\}
{\normalsize supervised by } 
\end{center}
\begin{center}
{\large Dr hab. Paweł Moskal}
\end{center}
\begin{center}
\end{center}
\begin{center}\end{center}
\begin{center}\end{center}
\begin{center}
\begin{figure}[h]
\hspace{6.7cm}
\vspace{-3.cm}
\parbox{0.10\textwidth}{\centerline{\epsfig{file=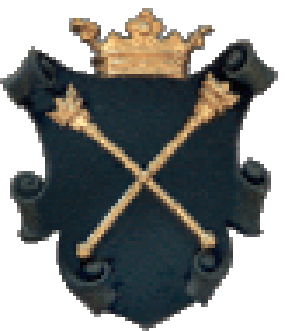,width=0.15\textwidth}}}
\end{figure}
\end{center}
\begin{center}
{\vspace{1.7cm}\normalsize KRAKÓW 2008}
\end{center}

\newpage
\thispagestyle{empty}
\begin{center}
\end{center}
\newpage
\thispagestyle{empty}
\begin{center}
{\bf\large Abstract}
\end{center}
\begin{center}
{\bf Analysis of the differential cross sections for the reaction $pp\rightarrow ppK^{+}K^{-}$ \\
in view of the $K^{+}K^{-}$ interaction}
\end{center}
Measurements of the $pp\rightarrow ppK^{+}K^{-}$ reaction, performed with the experiment COSY-11 
at the Cooler Synchrotron COSY, show a significant difference between the obtained excitation function 
and theoretical expectations including $pp$--FSI . The discrepancy may be assigned to the influence 
of $K^{+}K^{-}$ or $K^{-}p$ interaction. This interaction should manifest itself in the distributions
 of the differential cross section. This thesis presents an analysis of event distributions as a 
 function of the invariant masses of two particle subsystems. In particular in the analysis 
 two generalizations of the Dalitz plot proposed by Goldhaber and Nyborg are used. The present Investigations 
 are based on the experimental data determined by the COSY-11 collaboration from two measurements 
 at excess energies of Q~=~10~MeV and 28~MeV. The experimental distributions are compared to results 
 of Monte Carlo simulations generated with various parameters of the $K^{+}K^{-}$ and $K^{-}p$ interaction.\\
 The values of the $K^+K^-$ scattering length, extracted from two data sets for Q~=~10~MeV and 28~MeV amount to:
\begin{center}
$a_{K^+K^-}$~=~(11~$\pm$~8)~+~$i$(0~$\pm$~6)~fm for Q~=~10~MeV~,
\end{center}
and
\begin{center}
$a_{K^+K^-}$~=~(0.2~$\pm$~0.2)~+~$i$(0.0~$\pm$~0.5)~fm for Q~=~28~MeV~,
\end{center}
Due to the low statistics, the extracted values have large uncertainties and are consistent with very low values of the real 
and imaginary part of the scattering length.
 %
\newpage
\thispagestyle{empty}
     \normalsize
     \def\contentsname{\large \mbox{} \hspace{6cm} Contents}
     \tableofcontents
     \pagestyle{myheadings}
     \markboth{Contents}{Contents}
     \clearpage
     \normalsize
\fancyhead{}
\fancyfoot{}
\pagestyle{myheadings}
\chapter{Introduction}
\hspace{\parindent}
A primary goal of hadronic physics is to understand the mechanism of production and decays of mesons and
baryons and their structure in terms of quarks and gluons. However the non-perturbative
character of the Quantum Chromo Dynamics in the regime of low momentum transfer 
makes calculations impossible in terms of quark and gluon degrees of freedom.
~One of the most successful theory of strong forces at low energies is the 
lattice QCD, which is, however, not yet in the position to make quantitative statements about light scalar states ($J^{p}=0^{+}$). 
Therefore in this case one has to use other QCD based approaches, for example the constituent quark model. 
This approach treats the lightest scalar resonances $a_{0}(980)$ and $f_{0}(980)$ as a conventional $q\overline{q}$ states\cite{qq}. However, also this model cannot describe fully the properies of these particles and therefore
 they are considered as possible candidates for exotic four quark states ($qq\overline{q}\overline{q}$)\cite{qqqq}, quark-less gluonic hadrons\cite{gluon} or $K\overline{K}$ molecules\cite{hanhart,1,2}. Another candidate for 
a kaon bound state is the excited hyperon $\Lambda(1405)$, which is considered as a $\overline{K}N$ molecule\cite{KN}. For the formation
of both $K\overline{K}$ or $\overline{K}N$ molecules a crucial quantity is the strength of the kaon-antikaon or 
kaon-nucleon interaction, respectively. These interactions appear to be very important also with respect to many other 
physics phenomena, like for example a modification of the neutron star properties due to possible kaon 
condensation\cite{n-star,n-star1}. An Neutron  star consists mainly of neutrons, protons and leptons. However inside 
the core of the star the high nuclear density decreases the mas of $K^{-}$ allowing creation in processes 
like e.g. $e^{-}\rightarrow K^{-}\nu_{e}$ , and $n\rightarrow pK^{-}$. Since kaons are bosons, 
they can go into a Bose-Einstein condensate. Such phase transition
causes changes in neutron star parameters or may even trigger off its collapse into a black hole\cite{n-star2}. \\
In order to learn more about the above mentioned intriguing physical issues it is mandatory to conduct investigations 
of the $K\overline{K}$ and $\overline{K}N$ interaction. 
Because kaon targets are still unavailable, one of the realistic ways to study $K\overline{K}$ interaction
is the near threshold kaon pair production, for example in multi particle exit channels like
$pp\rightarrow ppK^{+}K^{-}$. Measurements of the total cross section of the aforementioned reaction
were performed at the cooler synchrotron COSY near the kinematical threshold\cite{cosy1,cosy2,wolke} by the COSY-11 
collaboration\cite{c-11}, and for higher energies by the ANKE collaboration\cite{anke}. Such an experiment
was conducted also by the DISTO collaboration\cite{disto} at Q~=~114 MeV at the SATURN accelerator.
The results indicate that the near threshold total cross section data points lie significantly 
above theoretical expectations, which neglet the interaction of $K^{+}K^{-}$ pair. \\
Encouraged by this observation, in order to deepen our knowledge about the low energy dynamics 
of the $ppK^{+}K^{-}$ system, we extended the analysis of the $pp\rightarrow ppK^{+}K^{-}$ reaction 
into the differential cross sections.  
This thesis concerns the investigation of event distributions for the 
above mentioned reaction measured by means of the COSY-11 detector setup at excess energies of Q~=~10~MeV and 
28~MeV\cite{cosy2}. The analysis is based 
on generalizations of the Dalitz plot for four particles proposed by Goldhaber\cite{goldhaber1,goldhaber2}
 and Nyborg\cite{nyborg}. \\
The next chapter of this thesis includes general information about kaon pair production in proton-proton collisions, 
and presents the main components of the COSY-11 apparatus\cite{c-11,smyr_c11,moskal_c11}. 
It contains also a brief description of measurements of the $pp\rightarrow ppK^{+}K^{-}$ 
reaction performed with this detection setup. In the third chapter we describe general properties of the Dalitz plot
which is a convenient tool in the investigation of the final state interaction for three particles\cite{grzonka}. 
Subsequently the next sections present two theoretical approaches  to  generalize the Dalitz plot for 
the case of the four particles in the final state.\\
The studies of the $K^{+}K^{-}$ final state interaction are described in Chapter 4. It presents 
the differential cross sections obtained from measurements at both excess energies and shortly describes 
the procedure that we applied to derive them from raw experimental data. The remaining part of the chapter 
is devoted to the determination of the scattering length of the $K^{+}K^{-}$ interaction. We present the main steps of the 
analysis and the obtained results.\\ 
The last chapter comprises summary and  perspectives of further studies of the $K^{+}K^{-}$ interaction. 
\pagestyle{myheadings}
\chapter{Kaon pair production with the COSY-11 experiment}
\hspace{\parindent}
\section{Near threshold kaon pair production in proton-proton collisions}
\hspace{\parindent}
There are several mechanisms that can lead to the production of kaon–antikaon pairs in 
nucleon–nucleon collisions. 
These can be divided mainly into three general classes\cite{anke}: \\
\\
a) direct production without any intermediate states:
 \begin{center}
$pp\rightarrow ppK^{+}K^{-}~,$ \\
\end{center}
b) the production via a meson, for example $\phi$, or at threshold - $a_{0}(980)$/$f_{0}(980)$, which then decays into $K\bar{K}$:
\begin{center}
$pp\rightarrow pp\phi \rightarrow ppK^{+}K^{-}~,$
\end{center}
\begin{center}
$pp\rightarrow ppa_{0}(980)/f_{0}(980)\rightarrow ppK^{+}K^{-}$
\end{center}
c) the associated production of $KY^{*}$, where the $\bar{K}$ is created through the decay of the excited hyperon $Y^{*}$:
 \begin{center}
$pp\rightarrow K^{+}pY^{*}\rightarrow ppK^{+}K^{-}$ \\
\end{center}
There are several excited hyperons that could contribute to such a process. Of particular interest 
 for low energy production are the $\Lambda(1405)$ and $\Sigma(1385)$.\\
 It is worth mentioning, that the mechanism of the near threshold $pp\rightarrow ppK^{+}K^{-}$ reaction is still unknown. 
It is therefore an interesting and still open question whether it is a direct reaction, or whether it should
be viewed as a two-step process through a mesonic state like $a_{0}$/$f_{0}$ 
or an excited hyperon $\Lambda(1405)/\Sigma(1385)$.\\
%

\section{COSY-11 detection setup}
\begin{center}
\begin{figure}[H] \parbox{0.4\textwidth}{\centerline{\epsfig{file=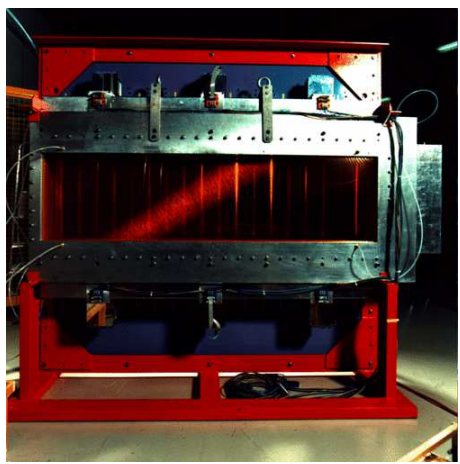,width=0.40\textwidth}}}
\caption{Front view of one of the COSY-11 drift chambers stack.
\label{dc1}
}
\end{figure}
\end{center}

\begin{center}
\begin{figure}[H] \parbox{0.5\textwidth}{\centerline{\epsfig{file=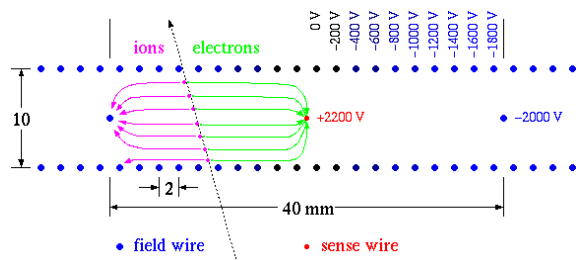,width=0.50\textwidth}}}
\caption{Schematic view of a drift chambers cell.
\label{dc2}
}
\end{figure}
\end{center}

\begin{center}
\begin{figure}[H] \parbox{0.4\textwidth}{\centerline{\epsfig{file=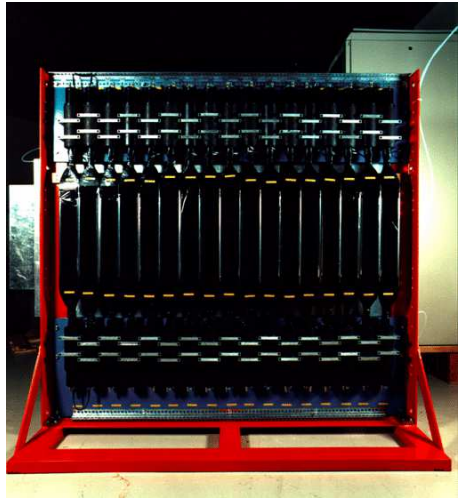,width=0.4\textwidth}}}
\caption{ Front view of the S1 hodoscope.
\label{S1}}
\end{figure}
\end{center}

\begin{center}
\begin{figure}[H] \parbox{0.4\textwidth}{\centerline{\epsfig{file=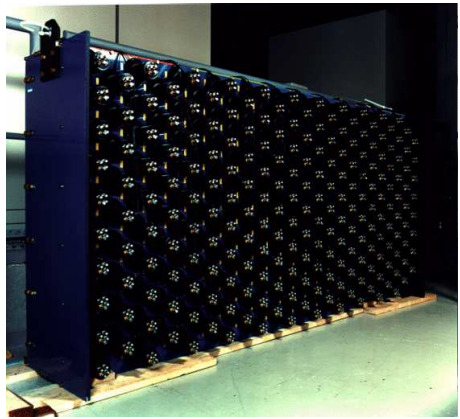,width=0.40\textwidth}}}
\caption{The COSY–11 large scintillator wall called S3 or ”AMADEUS”.
\label{S3}}
\end{figure}
\end{center}

\begin{center}
\begin{figure}[H] \parbox{0.35\textwidth}{\centerline{\epsfig{file=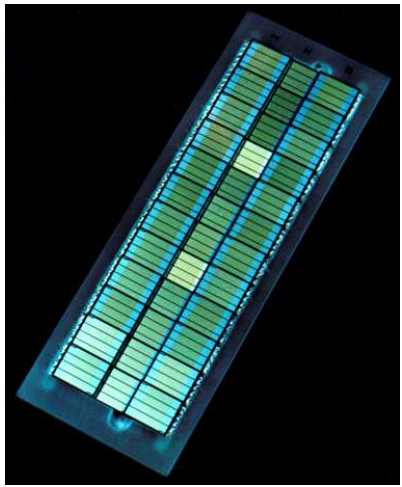,width=0.35\textwidth}}}
\caption{ One module of the COSY-11 silicon pad detector consisting of 144 independent detection units
arranged in three layers.
\label{silicon}}
\end{figure}
\end{center}

\hspace{\parindent}
Studies of the near threshold $pp\rightarrow ppK^{+}K^{-}$ reaction have been made possible due to 
the low emittance and small momentum spread proton beams available at the storage ring facilities and in
particular at the cooler synchrotron COSY\cite{prasuhn,maier} placed in Jülich in Germany. COSY provides polarized proton
 and deuteron beams with momenta up to $p~=~3.65$~GeV/c, corresponding to beam energies of $T_{proton}~=~2.83$
and $T_{deuteron}~=~2.23$~GeV, respectively. The beam can be cooled with an electron cooling system for low energies,
 or a stochastic cooling system for high energies. One of the detector systems operating over more then 
 eleven years at the COSY-ring was COSY-11. \\
The COSY-11 facility was designed for close-to-threshold reaction studies where the
relative momenta between the reaction particles are very small and all particles are focussed
into a narrow forward cone resulting in a high detection efficiency achievable by still using
rather small detection systems\cite{booklet}.
There is a special interest in the near threshold region because the final state interaction
between the outgoing reaction products is best visible and the contributing partial waves are
strongly limited mostly to pure s-waves which simplifies the theoretical description.
Due to the very strong energy dependence at threshold it is important to have a precise
knowledge of the beam momentum and a very good momentum resolution of the beam
particles\cite{pawel}.
\begin{figure}
\centering 
\includegraphics[angle=270,width=0.7\textwidth]{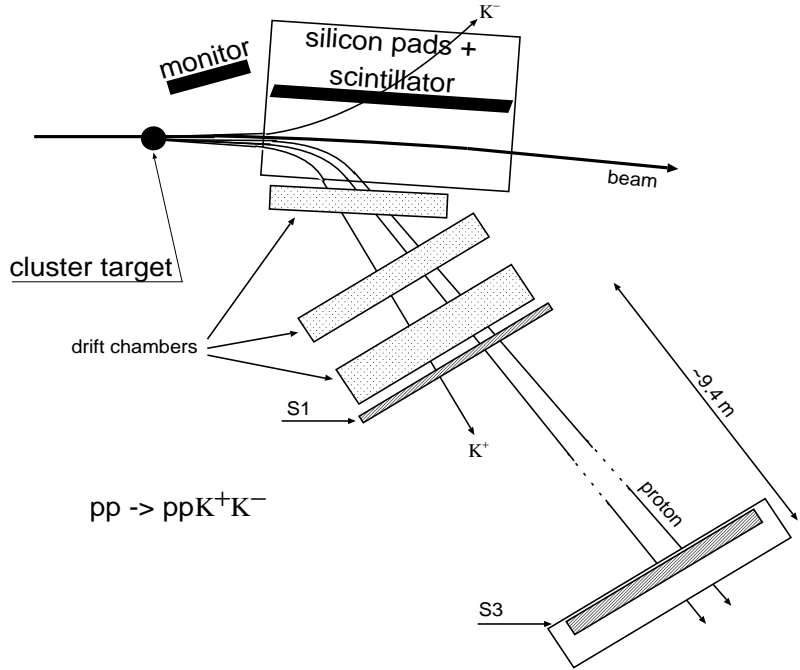}
\caption{Schematic view of the COSY-11 detector\cite{c-11,smyr_c11} with an exemplary event of the $pp\rightarrow ppK^{+}K^{-}$
 reaction channel. For the description see text.
 \label{c-11_rys}
 }
\end{figure}
COSY-11 is an internal magnetic spectrometer using a COSY machine dipole to separate
the charged reaction products from the circulating beam\cite{c-11}.
The detection system was already described in many publications\cite{booklet,c-11,cosy1,cosy2,wolke,habilitacja,moskal1},
therefore we will only briefly review its main components shown in Fig.~\ref{c-11_rys}. 
The description is based mainly on the references \cite{booklet,habilitacja,moskal1}. \\
The COSY-11 hydrogen cluster target~\cite{dombrowski} was installed in front of one of the regular COSY dipole magnets.
An internal target should be so thin that it does not affect the beam significantly, on the other hand it 
should be thick enough to reach a reasonable number of the required reactions. These requirements can be 
fulfilled with a cluster target where cooled gas (e.g. hydrogen) is pressed through 
a so-called 'Laval'-nozzle\cite{c-11,dombrowski}. 
Inside the nozzle the gas is cooled by the expansion and molecules freeze together forming clusters. 
Using skimmers free gas molecules are peeled off the cluster beam. So the rest gas pressure in the 
target chamber is very low. 
After traversing the target chamber the clusters hit the beam dump, evaporate, and are pumped away\cite{dombrowski}.  
The COSY-11 cluster target being a beam of $H_{2}$ molecules grouped to clusters of up to $10^{6}$ atoms crosses perpendicularly
the COSY beam with intensities of about $5 \cdot 10^{10}$ protons. The very thin target makes the 
probability of secondary scattering negligible and hence allows the precise determination of the ejectile momenta\cite{moskal1}. 
Moreover the energy loss of the COSY beam circulating through such a target could be compensated by the stochastic 
cooling system of COSY which guarantees a precise beam momentum over the whole measurement cycle. \\
 The main component of the COSY-11 detection system are two drift chambers~(see Fig.~\ref{dc1}) 
sets with 6 and 8 layers of vertical and inclined wires\cite{smyrski}. 
The drift chambers are used to determine the track of the charged particles. For this purpose thin wires 
are fixed in a volume filled with a special gas (in case of COSY-11 a mixture of Argon and Ethan) in a way, 
that the wires form cells as it is depicted in Fig.~\ref{dc2}~\cite{c-11}.  
Inside these cells a traversing charged particle ionizes the gas. Due to the electrical potentials applied to the wires the electrons drift to the sense wire and the connected electronics measures the time when the signal appears. The difference between this time and the time when the particle traversed the cell (measured by scintillation detectors) is used to reconstruct the impact point of the particle in the chambers midplane. 
The achieved position resolution of the COSY-11 drift chambers amount to about 200 $\mu m$\cite{booklet}.\\
To select candidates for desired events already on the trigger level, various scintillation detectors
were installed\cite{booklet}. Scintillation counters detect charged particles by the light which is produced
 when a particle travels through the scintillating material. The light is collected by light guides and
  converted into an electronic signal by photomultipliers\cite{wolke}. Behind the COSY-11 
drift chambers a hodoscope named S1 was mounted (Fig.~\ref{S1}). S1 consists of 16 thin scintillator bars with  
dimensions of 45 cm $\times $ 10 cm $\times$ 0.4 cm, each readout on both sides. 
This scintilator hodoscope measures the number of produced charged particles and serves as a start detector 
for a time-of-flight measurement. 
Nine meters apart from S1 a scintilator wall S3, read out by a matrix of 217 photomultipliers, was installed 
(see Fig.~\ref{S3}).
It serves as a position sensitive detector and as a stop counter for the time-of-flight measurement.
Because in the dipole gap there was no space for tracking detectors for negatively charged particles,
only a silicon pad detector combined with a scintillator was installed (see Fig.~\ref{silicon}). It determines 
the position of a charged particle within a binning of a few millimeters and
drastically reduces the background in the $K^{+}K^{-}$ production studies\cite{booklet}.
Other detector components are used for special purposes like a system of scintillator and
proportional wire chambers in forward direction above and below the vacuum chamber to
measure the luminosity and to determine the polarisation during measurements with polarized
beam\cite{booklet}.
A more detailed description of the detector components can be found for example in references
 \cite{c-11,smyr_c11,moskal_c11,booklet}.   
\section{Measurements of the $pp\rightarrow ppK^{+}K^{-}$ reaction at COSY-11 }
\begin{figure}
\centering 
\parbox{0.4\textwidth}{\centerline{\epsfig{file=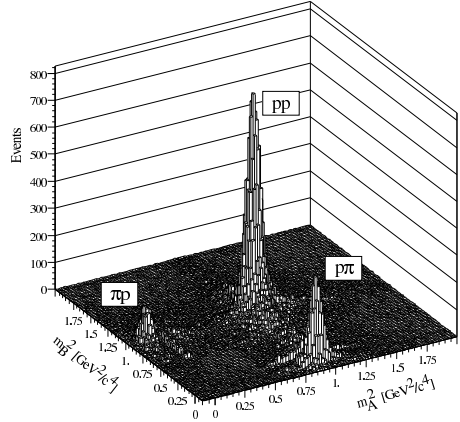,width=0.4\textwidth}}}
\caption{ Squared invariant masses of two positively charged particles
measured in coincidence at the right half of the drift chambers. The Figure is addapted from reference \cite{cosy2}.
\label{2proton}}
\end{figure}

\begin{figure}
\centering 
\parbox{0.4\textwidth}{\centerline{\epsfig{file=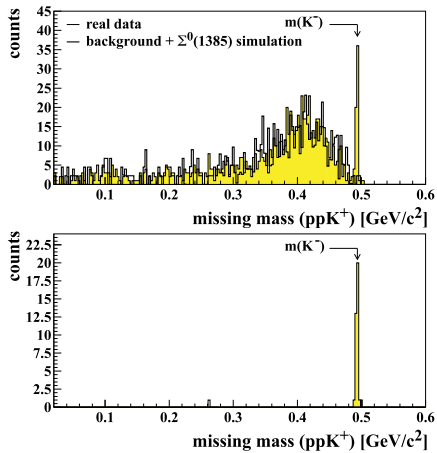,width=0.4\textwidth}}}
\caption{(\textbf{Upper panel}) Exemplary missing mass spectrum determined for the $pp\rightarrow ppK^{+}X^{-}$ reaction 
at an excess energy of Q = 17 MeV \cite{cosy1}; (\textbf{Lower panel}) The same spectrum with additional requirement 
of the signal in the dipol detector as it is described in the text. The Figure is addapted from refference \cite{cosy1}.
\label{mmass}}
\end{figure}

\hspace{\parindent}
The data used in our analysis, which will be presented in Chapter 4, were analysed and published in view of 
the total cross section\cite{cosy1}. Hereafter we will briefly describe the measurements method 
and for more details the interested reader is refered to articles \cite{booklet,moskal1}. 
The below presented description is based mostly on references\cite{c-11,moskal1}.\\
If at the intersection point of the cluster target and the COSY beam a collision of protons leads
to the production of a $K^{+}K^{-}$ meson pair, then the 
reaction products – having smaller
momenta than the circulating beam – are directed by the magnetic field towards the COSY-11
detection system and leave the vacuum chamber through the thin exit foils\cite{c-11}. Tracks
of the positively charged particles, registered by the drift chambers, are traced back
through the magnetic field to the nominal interaction point leading to a momentum 
determination. A simultaneous measurement of the velocity, performed by means of
scintillation detectors S1 and S3, permits to identify the registered particle and to determine
its four momentum vector. Since at threshold the center-of-mass momenta of the
produced particles are small compared to the beam momentum, in the laboratory frame
all ejectiles are moving with almost the same velocity. This means that the laboratory
proton momenta are almost two times larger then the momenta of kaons. Therefore,
in the dipole magnetic field protons have a much larger bending radius than kaons.
As a consequence, in case of the near threshold production, protons and kaons are
registered in separate parts of the drift chambers as shown schematically in Fig.~\ref{c-11_rys} 
\cite{moskal1}.
Fig.~\ref{2proton} shows the squared invariant mass of two simultaneously detected particles in the
near beampipe part of the drift chamber. A clear separation is observed into groups of events with
two protons and proton and pion\cite{moskal1}.
 This spectrum allows to select events with two registered protons. The additional requirement that
the mass of the third particle, registered at the far beampipe side of the chamber, corresponds
to the mass of the kaon, makes possible to identify events of 
a $pp\rightarrow ppK^{+}X^{-}$ reaction\cite{moskal1}.
Knowing both the four momenta of positively charged ejectiles and the
proton beam momentum one can calculate the mass of the unobserved system $X^{-}$ and 
its four momentum.
This method of particle identification is called the \emph{missing mass technique}\cite{maglic185}.
Fig.~\ref{mmass} (upper panel) presents an example of the missing mass spectrum with respect to
the identified $ppK^{+}$ subsystem.
 In case of the $pp\rightarrow ppK^{+}K^{-}$ reaction this should correspond to the mass of the $K^{-}$ meson, and indeed a pronounced signal can be clearly seen. 
 The additional broad structure seen in the Figure is partly due to the 
$pp\rightarrow pp\pi^{+}X^{-}$ reaction, where the $\pi^{+}$ was misidentified 
as a $K^{+}$ meson and partly due to the $K^{+}$ meson production 
associated with the hyperons $\Lambda$(1405) or $\Sigma$(1385), e.g. via
the reaction $pp\rightarrow pK^{+}\Lambda(1405)\rightarrow pK^{+}\Sigma \pi^{0}\rightarrow pK^{+}\Lambda \gamma \pi^{0}
\rightarrow ppK^{+} \pi^{-} \gamma \pi^{0}$\cite{cosy1,cosy2,wolke,moskal1}. In the latter case 
the missing mass of the $ppK^{+}$ system corresponds to the invariant mass of the $\pi^{+}\pi^{0}\gamma$
system and hence can acquire values from twice the pion mass up to the kinematical
limit. The background, however, can be completely reduced by demanding a signal
in the silicon strip detectors at the position where the $K^{-}$ meson originating from the
$pp\rightarrow ppK^{+}K^{-}$ reaction is expected\cite{moskal1} (the spectrum in the lower panel of Fig.~\ref{mmass}).
This clear identification allows to select events originating from the $pp\rightarrow ppK^{+}K^{-}$ reaction 
and to determine the total and differential cross section.\\
The results of all measurements of the $pp\rightarrow ppK^{+}K^{-}$ reaction performed with COSY-11 will be presented 
in the next section.
\section{Excitation function for the $pp\rightarrow ppK^{+}K^{-}$ reaction at threshold }
\hspace{\parindent}
The measurements of the $pp\rightarrow ppK^{+}K^{-}$ reaction were performed by the COSY-11 group several times 
in the period from 1995 to 2005\cite{cosy1,cosy2,wolke}. The results are presented together with data obtained by other 
collaborations in Fig.~\ref{excitation-f}: 
\begin{center}
\begin{figure}[H] \parbox{0.7\textwidth}{\centerline{\epsfig{file=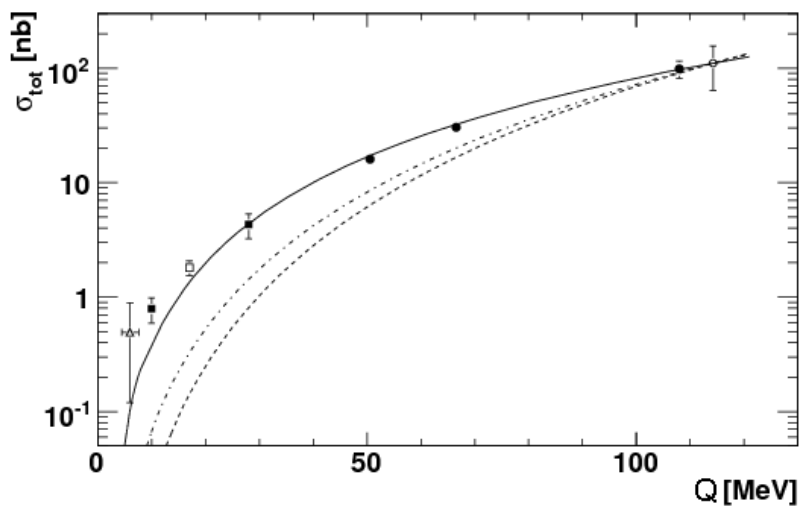,width=0.7\textwidth}}}
\caption{Total cross section as a function of the excess energy Q for the reaction $pp\rightarrow ppK^{+}K^{-}$. 
Open and filled circles represent the DISTO\cite{disto}(open circle) and ANKE\cite{anke} measurements respectively.
The four points nearest threshold are the results from COSY-11 measurements\cite{cosy1,cosy2,wolke}. The curves
are described in the text. The Figure is adapted from ref.\cite{anke}. 
\label{excitation-f}
}
\end{figure}
\end{center}
In the Figure one can see also three curves representing the theoretical expectations. The dashed curve represents 
the energy dependence from four-body phase space, when we assume that there is no interaction between 
particles in the final state. This result differs from the experimental data by two orders of magnitude 
at Q~=~10~MeV and by a factor of about five at Q~=~28~MeV. Hence, it is obvious, that the final state interactions   
effects in the $ppK^{+}K^{-}$ system cannot be neglected\cite{oelert}. 
Inclusion of the $pp$--FSI( dashed-dotted line in Fig.~\ref{excitation-f}) by folding its parameterization known 
from the three body final state with the four body phase space is clearly closer to the experimental
 results but does not fully account for the difference\cite{cosy2}. The enhancement may be due to the 
 influence of $K^{+}K^{-}$ or $Kp$ interaction which was neglected in the calculations. 
 Indeed, the inclusion of the $K^{-}p$--FSI (solid line) reproduces the experimental data for the exess energies 
 down to Q~=~28~MeV. This model neglects the $K^{+}p$ interaction, since it is repulsive and rather 
 weak\cite{moskal1,balewski}, and hence it can be interpreted as an extra attraction in the $K^{-}p$ system 
 \cite{anke}. However, the data very near threshold still remain underestimated, which indicates that in this region 
the influence of the $K^{+}K^{-}$ interaction may be significant and cannot be neglected. \\
 The interaction may manifest itself even stronger in the distributions of the 
differential cross sections\cite{pawel}. A significant effect observed for the excitation function
for the $pp\rightarrow ppK^{+}K^{-}$ reaction encouraged us to carry out 
the analysis in spite of the fact that the available statistics is quite low.
The investigations of the $K^{+}K^{-}$ interaction described in this thesis is based on experimental
data obtained from two COSY-11 measurements at excess energies of Q~=~10~MeV (27 events)
and 28~MeV (30 events)\cite{cosy2}.
%
\pagestyle{myheadings}
\chapter{Dalitz plot and its generalization}
\hspace{\parindent}
\section{The Dalitz plot as a tool in particle physics}
\begin{figure}[H]
\centering 
\parbox{0.4\textwidth}{\centerline{\epsfig{file=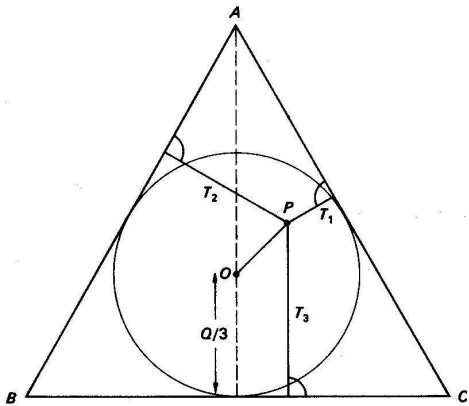,width=0.4\textwidth}}}
\caption{Exemplary event in the Dalitz plot representation. Q = $T_{1}+T_{2}+T_{3}$ denotes the total 
kinetic energy of the three particle system.
\label{triangle}
}
\end{figure}

\hspace{\parindent}
In many investigations of elementary particle reactions the distributions of energies
and momenta of particles in the final state are measured. The measured distributions are then 
compared with the theoretical expectations 
 in order to extract parameters expressing final state interaction between particles\cite{nyborg}.
For three particles in the final state the process of analysing
data by plotting them in the space of two internal parameters is well known. It was originated
by Dalitz in a nonrelativistic application. In the original article\cite{dalitz} Dalitz let the
distances to the sides of an equilateral triangle be the energies of the three particles in
the centre-of-mass frame. The sum of distances from a point within the triangle to its
sides is constant and equal to the height, which represents the total energy of the system (Fig.~\ref{triangle}).
 In the nonrelativistic approximation for particles with equal masses the physical allowed 
 region on the Dalitz plot is  bounded by a circle inscribed in the triangle
 due to momentum conservation (Fig.~\ref{triangle})\cite{perkins}. Therefore,
the interior points fulfil 
 four-momentum conservation and represent energy partitions.
The relativistic extension of Dalitz’s analysis was given by Fabri\cite{fabri}.
\begin{figure}
\centering 
\parbox{0.45\textwidth}{\centerline{\epsfig{file=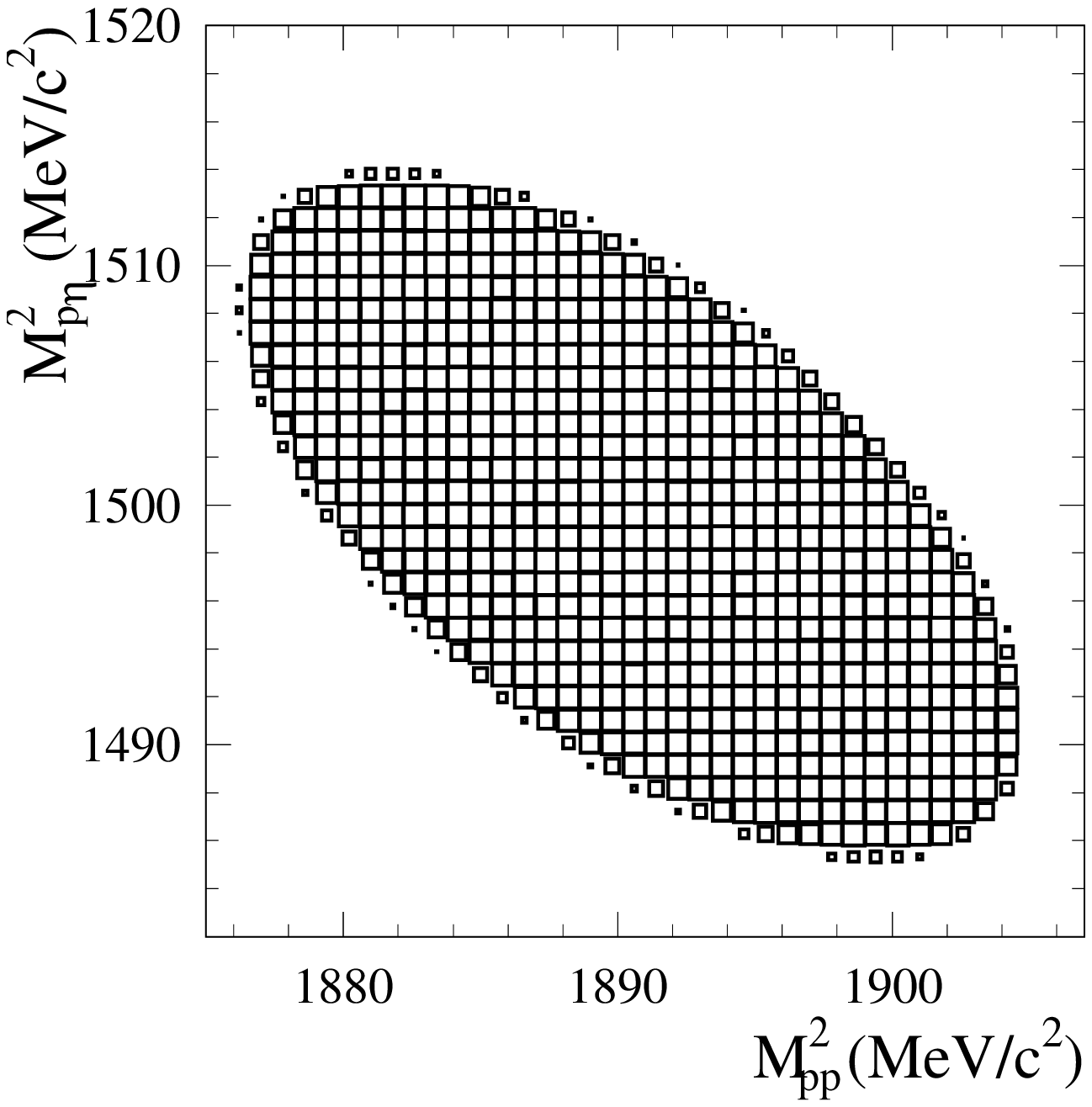,width=0.4\textwidth}}}
\parbox{0.45\textwidth}{\centerline{\epsfig{file=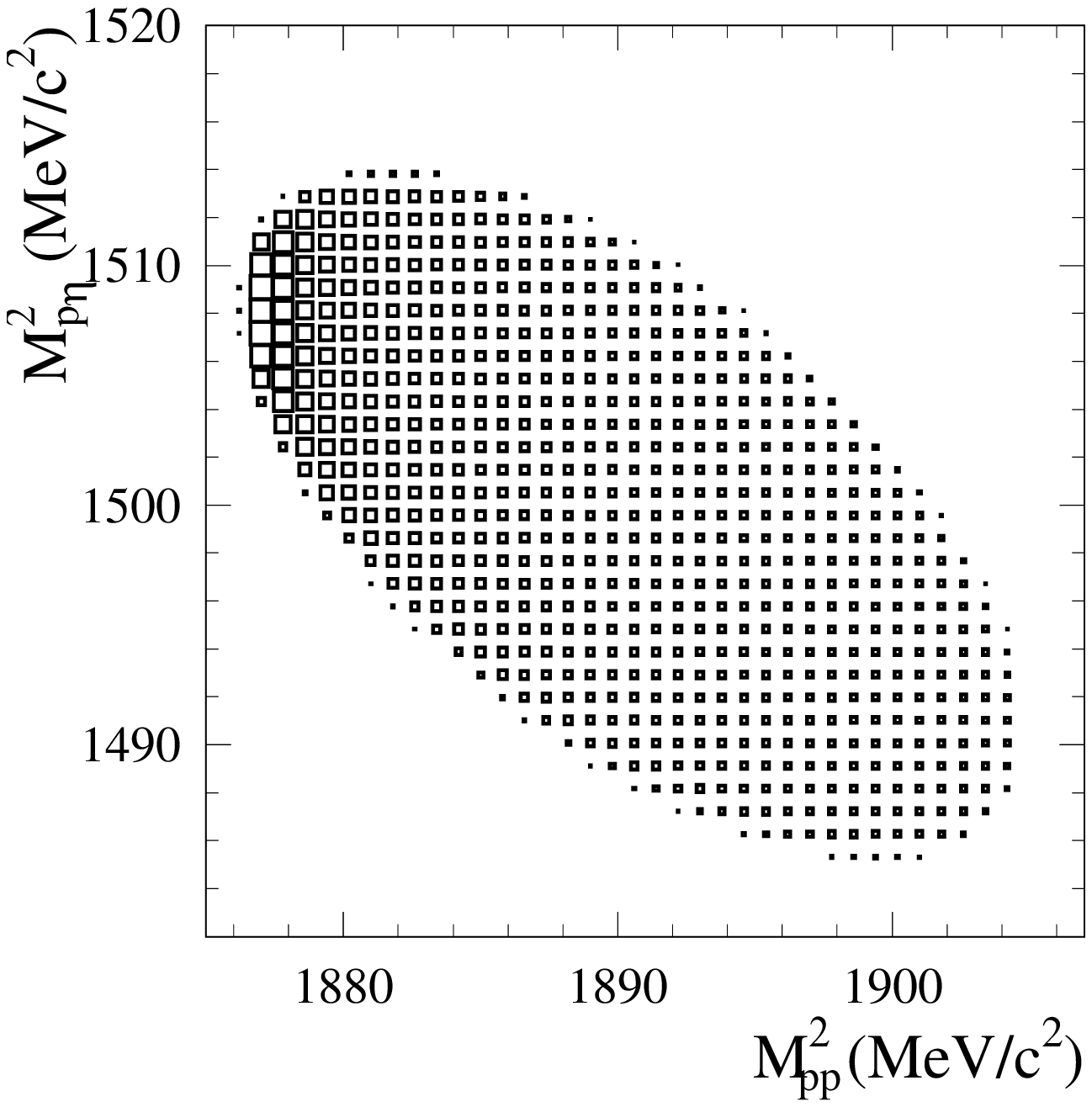,width=0.4\textwidth}}}
\caption{Monte Carlo Dalitz plot distribution for the $pp\rightarrow pp\eta$ reaction at Q~=~28~MeV for 
the homogeneously populated phase space (left picture), and its modification by proton-proton final
 state interaction (right picture). The areas of the squares are proportional to the number 
of entries and are shown in a linear scale.}
\label{dalitz_wyk}
\end{figure}
The interaction between particles depends on their relative momenta, thus for investigations of final state interactions 
more convenient variables then energies are the squared invariant masses of the two-body subsystems\cite{grzonka}.
 The squared invariant mass $M_{ij}^{2}$ of particles $i$ and $j$ is defined as:
\begin{equation}
M_{ij}^{2}~=~\left(E_{i}+E_{j}\right)^{2}-\left(\vec{p_{i}}+\vec{p_{j}}\right)^{2}  
\label{invariant_mass}
\end{equation}
where $E_{i}, E_{j},\vec{p_{i}},\vec{p_{j}}$ denote the energies and momenta.
In a geometrical representation the Dalitz plot is a plane in space of three squared invariant masses -
 ($M_{12}^{2},M_{23}^{2},M_{13}^{2}$),  
which is orthogonal to the space diagonal\cite{grzonka}.
  From the three invariant masses only two are independent due to the relation\cite{grzonka}:
\begin{equation}
M_{12}^{2}+M_{13}^{2}+M_{23}^{2} = s+ m_{1}^{2}+m_{2}^{2}+m_{3}^{2}
\label{rownanie_inv_mass}
\end{equation}
 where $s$ denotes the square of the total energy of the system and $m_{i}$ is a mass of $i$-th particle. 
 Thus a projection of the physical region on any of the ($M_{ij}^{2}+M_{jk}^{2}$) planes still comprises the whole accessible information 
 about the three particle system. Such two dimensional event distributions are bounded by well defined
smooth closed curves and are lorentz invariant. For example the boundary of the projection on the ($M_{12}^{2},M_{23}^{2}$) plane is given 
by a following equation\cite{nyborg}:
\begin{equation}
G\left(M_{12}^{2},M_{23}^{2},m_{1}^{2},m_{3}^{2},\sqrt{s},m_{2}^{2}\right) = 0~,
\label{G_dalitz}
\end{equation}
where $G$ is defined as:
\begin{equation}
G\left(x,y,z,u,v,w\right) = x^{2}y+xy^{2}+z^{2}u+zu^{2}+v^{2}w+vw^{2}+xzw+xuv+yzv+yuw
\label{G_dalitz1}
\nonumber
\end{equation}
\begin{equation}
-xy\left(z+u+v+w\right)-zu\left(x+y+v+w\right)-vw\left(x+y+z+u\right)~.
\label{G_dalitz2}
\end{equation}
The Dalitz plot representation allows also for a simple interpretation
of the kinematically available phase space volume as an area of the plot\cite{habilitacja}:
\begin{equation}
V_{ps} = \int dV_{ps} = 
 \frac{\pi^{2}}{4\,\mbox{s}}
 \int\limits_{(\mbox{\scriptsize m}_1+\mbox{\scriptsize m}_2)^2}^{
   (\sqrt{\mbox{\scriptsize s}}-\mbox{\scriptsize m}_3)^2} d\,\mbox{M}_{12}^{2}
 \int\limits_{\mbox{\scriptsize M}_{23}^{2~min}(\mbox{\scriptsize M}_{12}^{2})}^{
   \mbox{\scriptsize M}_{23}^{2~max}(\mbox{\scriptsize M}_{12}^{2})} 
   d\,\mbox{M}_{23}^{2}\; = 
   \nonumber
\end{equation}
\begin{equation}
\label{Vps_relativistic}
=\;\frac{\pi^{2}}{4\,\mbox{s}}
 \int\limits_{(\mbox{\scriptsize m}_1+\mbox{\scriptsize m}_2)^2}^{
  (\sqrt{\mbox{\scriptsize s}}-\mbox{\scriptsize m}_3)^2}\;
 \frac{d\,\mbox{M}_{12}^{2}}{\mbox{M}_{12}^{2}}
 \sqrt{\lambda(\mbox{M}_{12}^{2},\mbox{s},\mbox{m}_3^2) 
       \lambda(\mbox{M}_{12}^{2},\mbox{m}_2^2,\mbox{m}_1^2)}, 
\end{equation}
where $\lambda$ is the kinematical triangle function defined as\cite{bycklingkajantie}:
\begin{equation}
\label{kaellen}
\lambda(x, y, z) = x^2 + y^2 + z^2 - 2xy - 2yz - 2zx.
\end{equation}
In case of no dynamics and the absence of any final state interaction between particles
the occupation of the Dalitz plot would be fully homogeneous because the creation in
any phase space interval would be equally probable\cite{habilitacja,grzonka}. Thus, final state interaction should
show up as a modification of the event density on the Dalitz plot. Such effect was observed
experimentally eg. by the COSY-11 collaboration for the $pp\rightarrow pp\eta$ reaction \cite{prc2004}.\\
In order to illustrate the effect in Fig.~\ref{dalitz_wyk} we show an example of how the 
uniformly populated phase space (left plot) is modified by the S-wave proton-proton 
final state interaction (right plot). It is clearly seen, that the interaction increases significantly 
the event density in the region where the relative momenta of the protons are small.\\
Thus, such event distributions are very convenient in analysis of the final state interaction 
and they are widely applied in practice. 

\section{Generalization of the Dalitz plot for the case of four particles}
\hspace{\parindent}
One can ask a question whether the analysis described in the last section can be made for four 
or more particles in the final state. The answer is affirmative, accorging to Nyborg the Dalitz 
plot can be generalized, under some assumptions, for any number of particles\cite{nyborg1}. 
In this thesis we present only three from many possible generalizations. In particular we will
introduce generalizations of the Dalitz plot for 
four particles made by Chodrow\cite{chodrow}, Goldhaber\cite{goldhaber1,goldhaber2} and
Nyborg\cite{nyborg}. 
Further on the Goldhaber and Nyborg approaches will be used in our analysis of the $K^{+}K^{-}$ final state interaction.\\

\subsection{Chodrow plot}
\hspace{\parindent}
Consider a reaction like $a+b\longrightarrow 1+2+3+4$ yelding four particles with mases $m_{i}$ and total energy $\sqrt{s}$ in centre-of-mass frame.
The probability, that the momentum of the $i^{th}$ particle will be in a range $d^{3}p_{i}$ is given by:
\begin{equation}
d^{12}P=d^{3}p_{1}d^{3}p_{2}d^{3}p_{3}d^{3}p_{4}\frac{1}{16E_{1}E_{2}E_{3}E_{4}} \delta^{3}\left(\sum_{j=1}^4 \vec{p}_{j}\right)\delta\left(\sum_{j=1}^4 E_{j}-\sqrt{s} \right)\left|M\right|^{2}
\label{prawdopodob}
\end{equation}
where $E_{i}=\sqrt{\vec{p}^{2}_{i}+m^{2}_{i}}$ denotes an energy of the $i^{th}$
particle ($c$ = 1) and $M$ denotes the invariant matrix element for the process. 
In his work Chodrow assumed, that $M$ depends only on energies of the particles in the 
final state\cite{chodrow}. Under this assumption one can integrate the distribution expressed in 
eq.~\eqref{prawdopodob} over delta functions and over spatial orientations of the entire four body system. 
The resulting density of events in $\left(E_{1},E_{2},E_{3}\right)$ space is then given by\cite{chodrow}:
\begin{equation}
d^{3}P=dE_{1}dE_{2}dE_{3}\left|M\right|^{2}min\left(\left|\vec{p}_{1}\right|,\left|\vec{p}_{2}\right|\right)~.
\label{chodrow1}
\end{equation}
It is therefore possible to analyse resonances by plotting the event distribution 
in an $E_{1}E_{2}$-plane. However, the analysis is not easy due to the factor 
$min\left(\left|\vec{p}_{1}\right|,\left|\vec{p}_{2}\right|\right)~$\cite{chodrow}. 
This difficulty can be avoided if in the final state particles 1 and 2 are identical, like for example 
in the case of the $ppK^{+}K^{-}$ system. In such case the function $min\left(\left|\vec{p}_{1}\right|
,\left|\vec{p}_{2}\right|\right)$ becomes 
symetric and the analysis can be confined to the region of the $E_{1}E_{2}$-plane defined by condition $E_{1}<E_{2}$. 
Consequently from eq.~\eqref{chodrow1} one obtains\cite{chodrow}:
\begin{equation}
d^{3}P=32\pi^{2}\left|M\right|^{2}\sqrt{E_{1}^{2}-m_{1}^{2}}~dE_{1}dE_{2}dE_{3}~,
\label{chodrow2}
\end{equation}
or
\begin{equation}
d^{3}P=32\pi^{2}\left|M\right|^{2}dF_{1}dE_{2}dE_{3}
\label{chodrow3}
\end{equation}
where $dF_{1}=\sqrt{E^{2}_{1}-m^{2}_{1}}~dE_{1}$. 
This implies, after integration, that: 
\begin{equation}
F_{1}=\frac{1}{2}\left[E_{1}\sqrt{E^{2}_{1}-m^{2}_{1}}- m^{2}_{1}~cosh^{-1}\left(\frac{E_{1}}{m_{1}}\right)\right]~.
\label{chodrow4}
\end{equation} 
The distribution of events can then be plotted in the $F_{1}E_{2}$-plane and resonances may be directly read off the plot, 
like in case of three particles.
\begin{figure}
\centering 
\parbox{0.45\textwidth}{\centerline{\epsfig{file=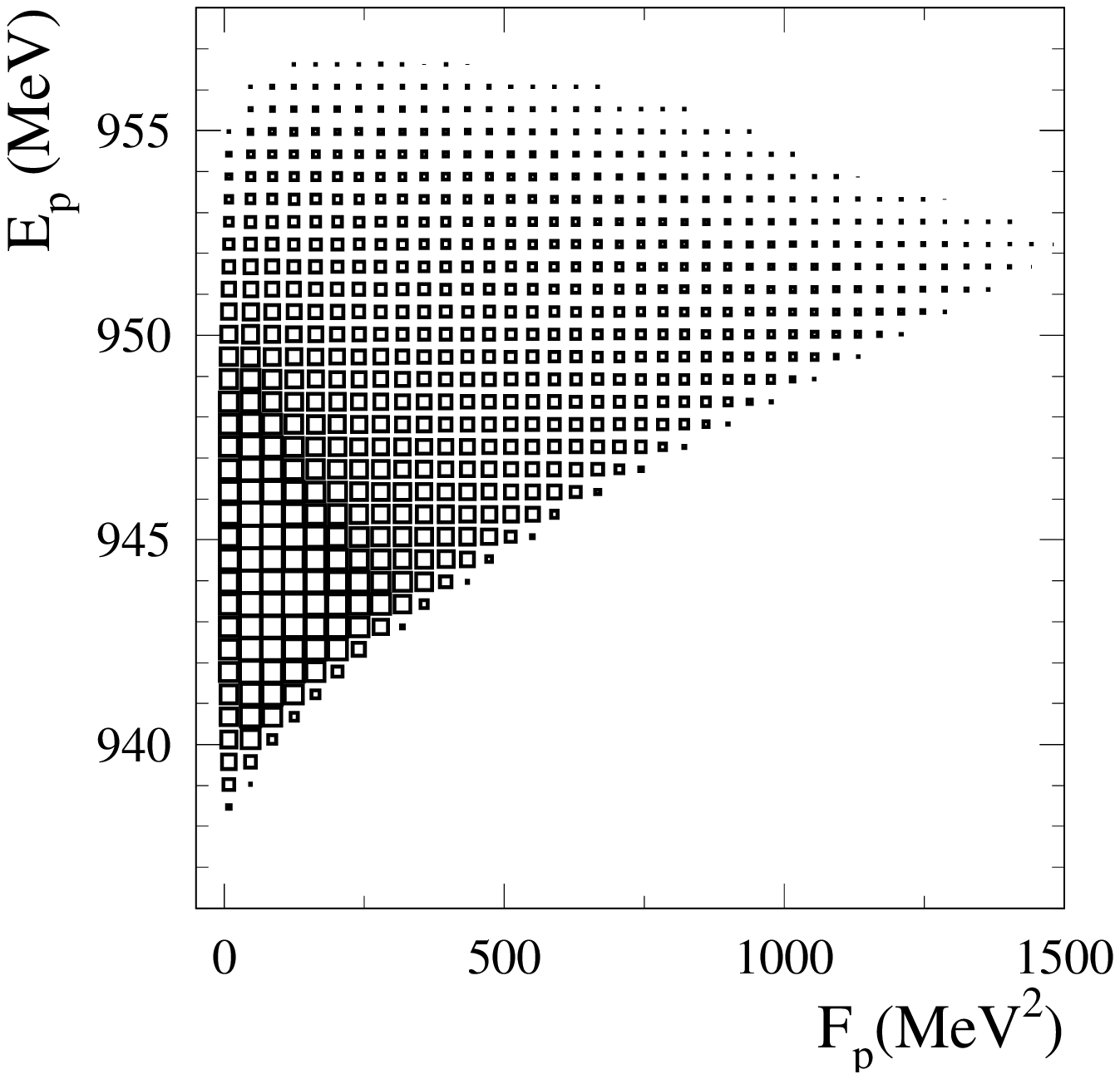,width=0.4\textwidth}}}
\parbox{0.45\textwidth}{\centerline{\epsfig{file=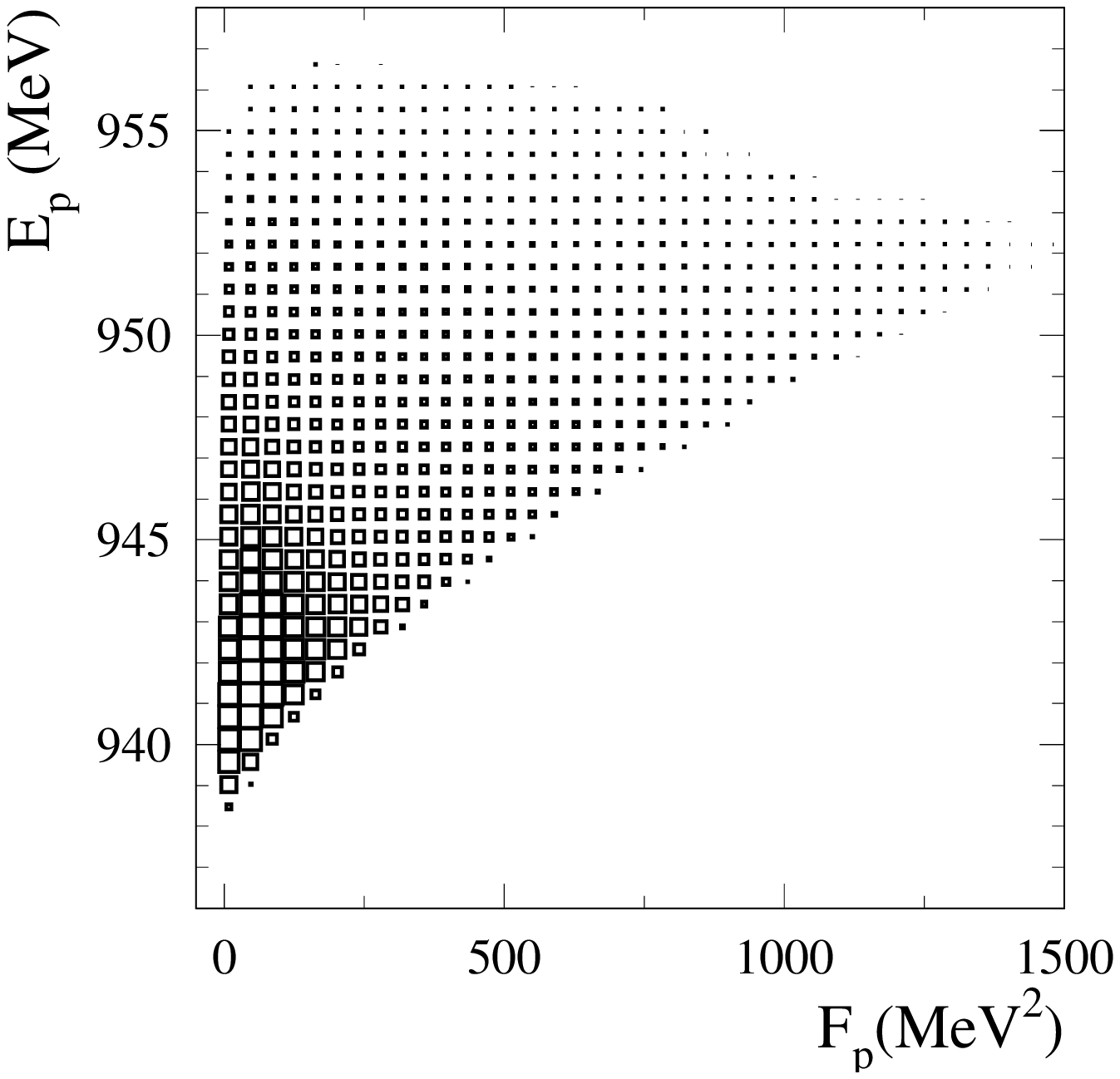,width=0.4\textwidth}}}
\caption{Chodrow plot for the $pp\rightarrow ppK^{+}K^{-}$ reaction simulated at Q~=~28~MeV for 
the homogeneously populated phase space (left plot), and for the case when proton-proton final
 state interaction is additionally taken into account (right plot).}
\label{chodrow_wyk}
\end{figure}
Fig.~\ref{chodrow_wyk} shows an exemplary Chodrow plot for the $pp\rightarrow ppK^{+}K^{-}$ 
reaction at Q~=~28~MeV simulated assuming homogeneous phase space distribution. The modification of the event 
density on the plot caused by proton-proton final state interaction ($pp$--FSI) is clearly seen by comparing
left and right plots.
 One can see that the physically allowed region on the Chodrow plot is bounded by a well defined curve, but the distribution 
 is not homogeneous which is obviously a disadvantage. It is also worth mentioning, that the event distribution in 
 $\left(F_{1},E_{2}\right)$ variables is not lorentz invariant, so one cannot compare Chodrow plots 
 determined in different reference frames\cite{kamys}.
 \subsection{Goldhaber and Nyborg plots}
\hspace{\parindent}
According to Nyborg  several other extensions of Dalitz plot can be obtained 
if one assumes, that the matrix element $M$ depends only on the invariant masses
 of two- and three particle subsytems\cite{nyborg}.
\begin{figure}[H]
\centering 
{\epsfig{file=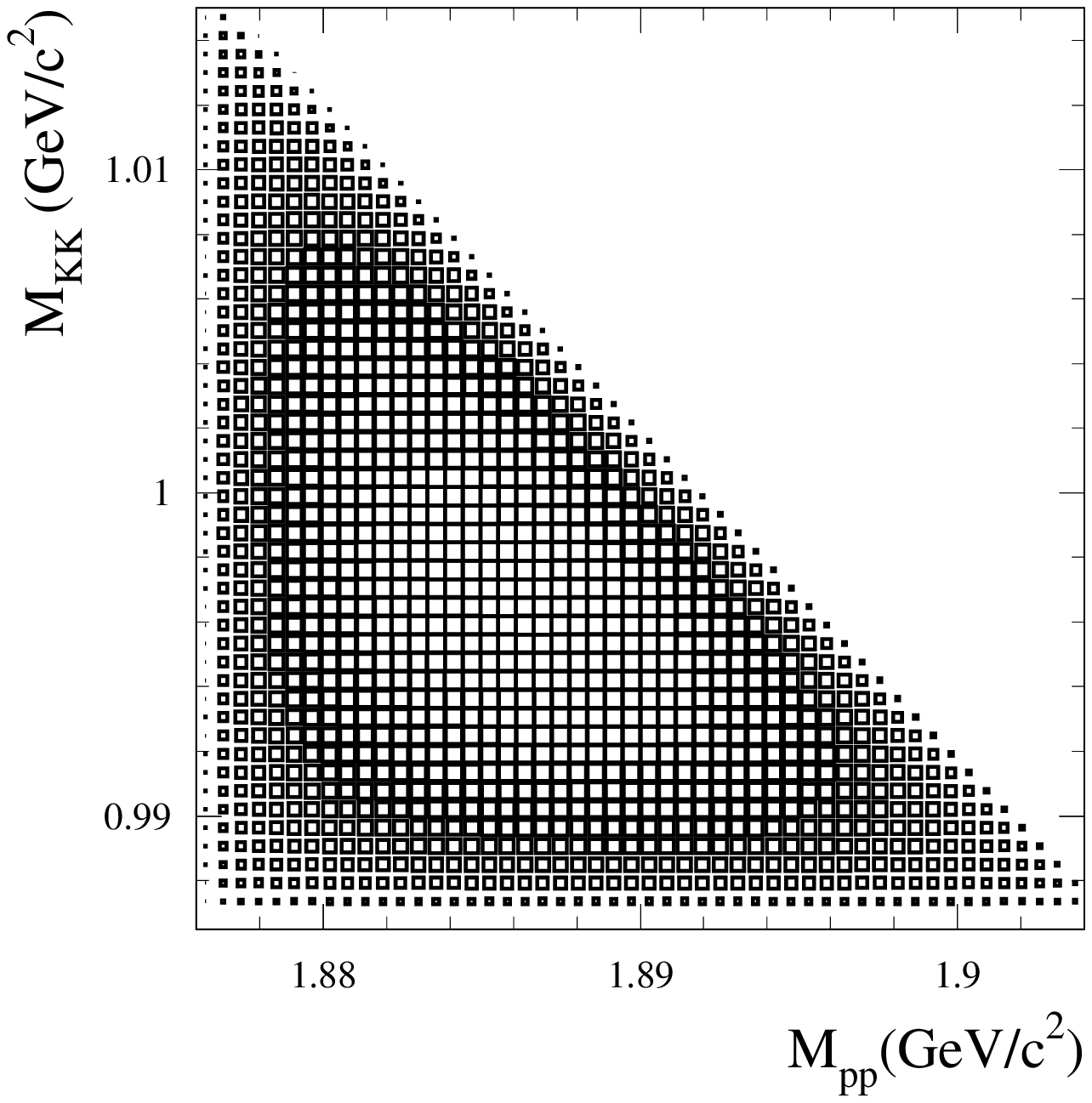,width=0.35\textwidth}}
{\epsfig{file=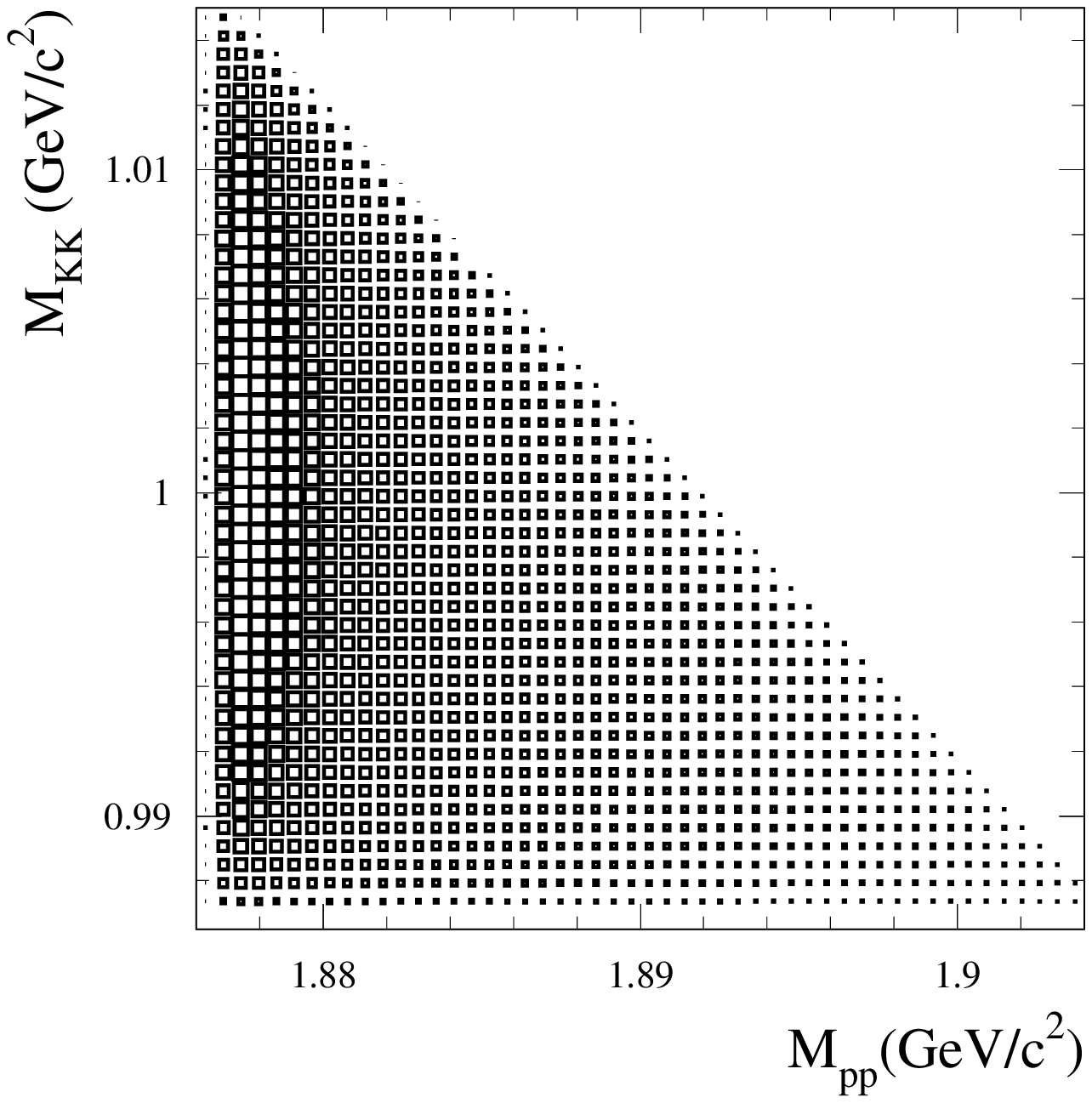,width=0.35\textwidth}}
{\epsfig{file=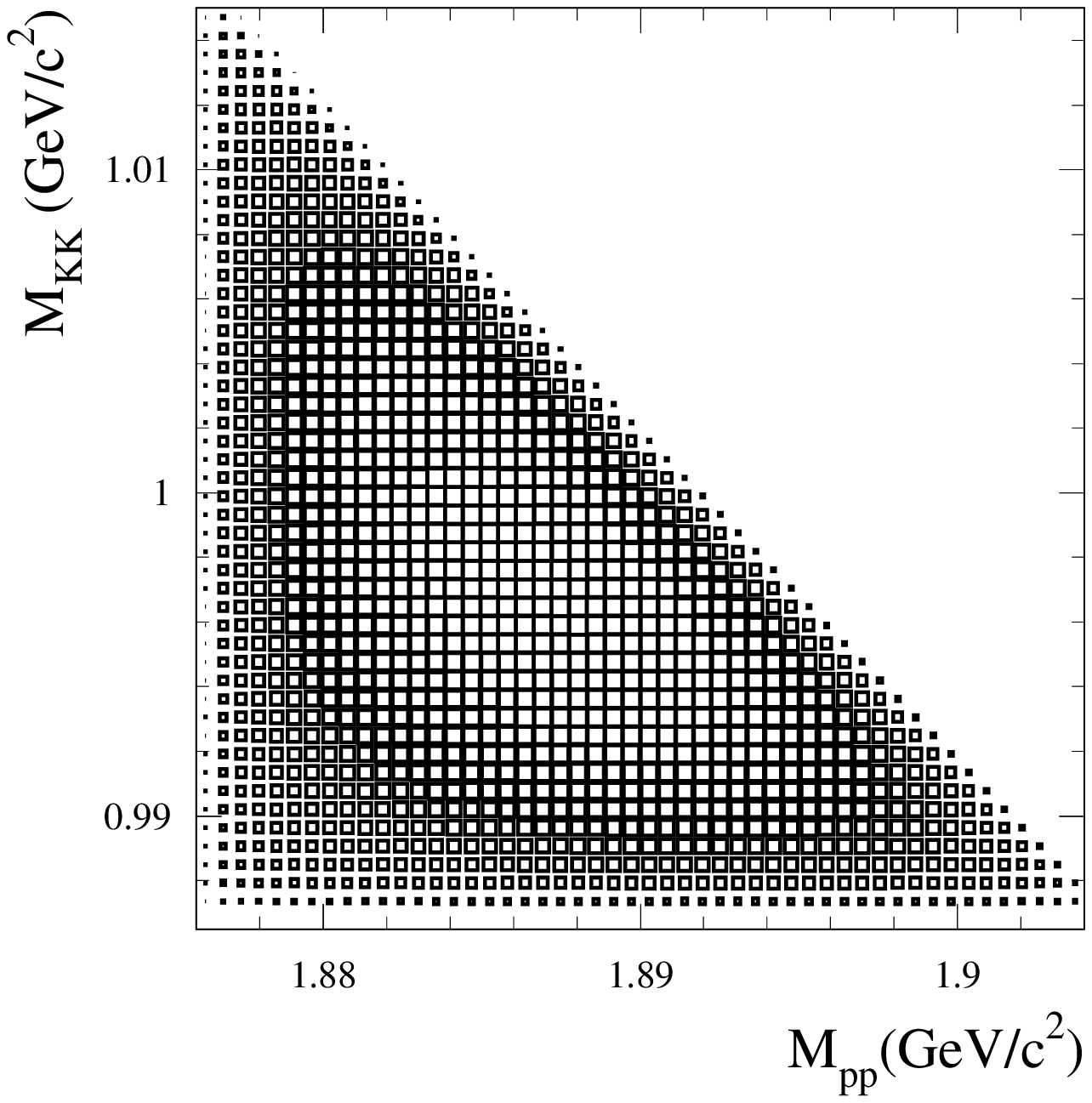,width=0.35\textwidth}}
\caption{Goldhaber plot for the $pp\rightarrow ppK^{+}K^{-}$ reaction simulated at Q~=~28~MeV for 
the homogeneously populated phase space (left plot), for the case when proton-proton final
 state interaction is additionally taken into account (middle plot), and with only $pK^{-}$ final state interaction included (right plot).}
\label{goldhaber_wyk}
\end{figure}
\begin{figure}[H]
\centering 
{\epsfig{file=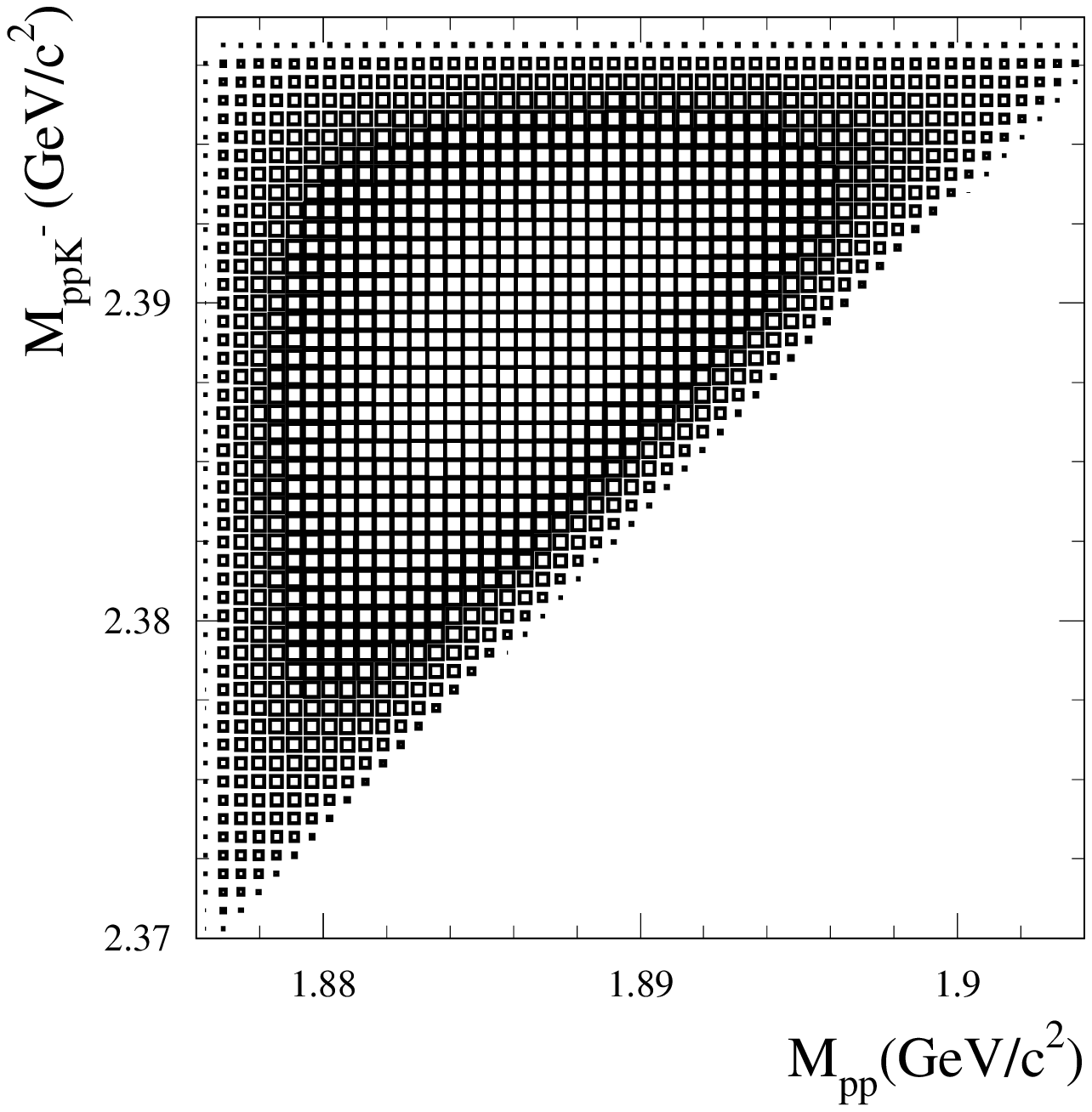,width=0.35\textwidth}}
{\epsfig{file=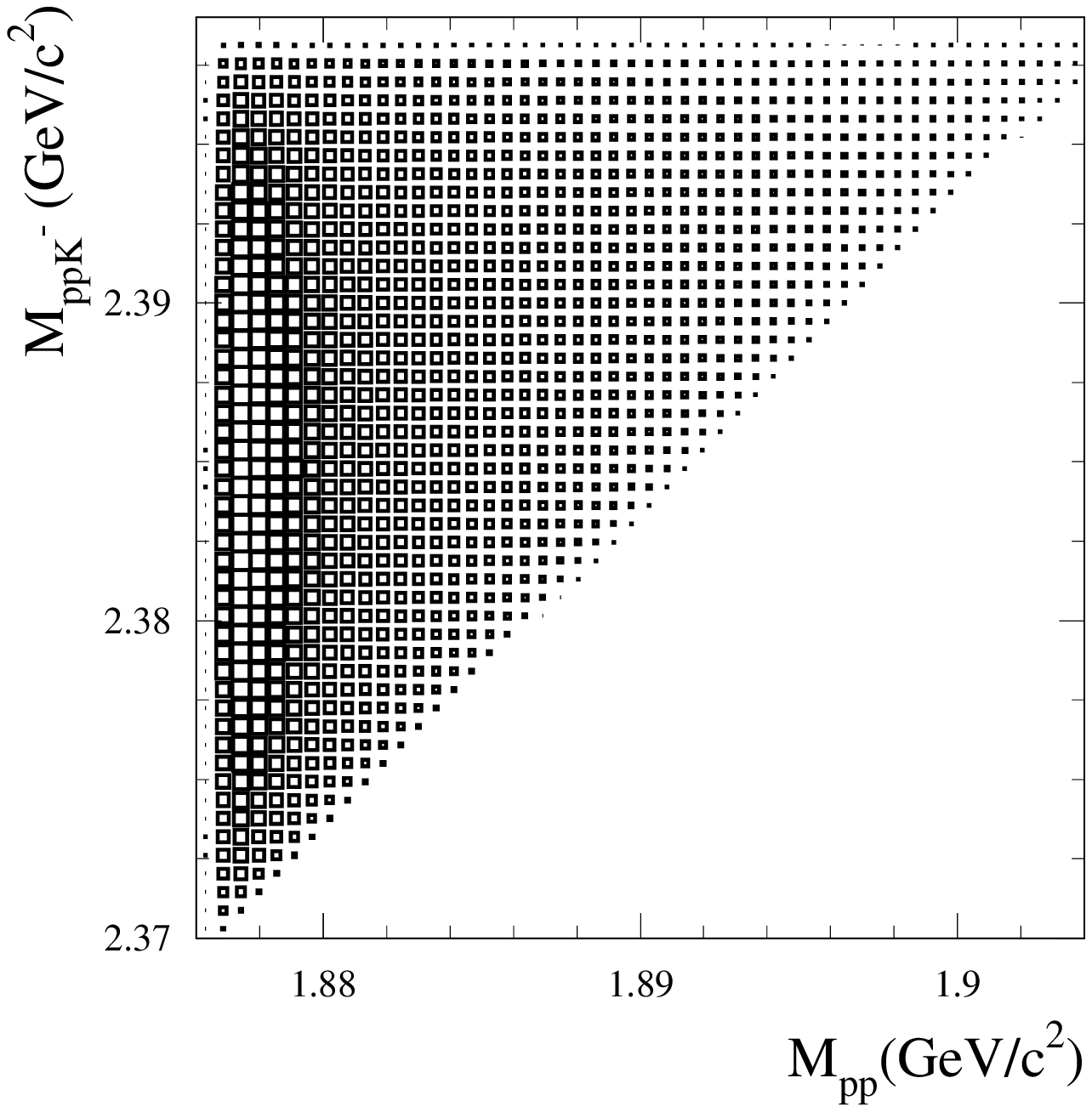,width=0.35\textwidth}}
{\epsfig{file=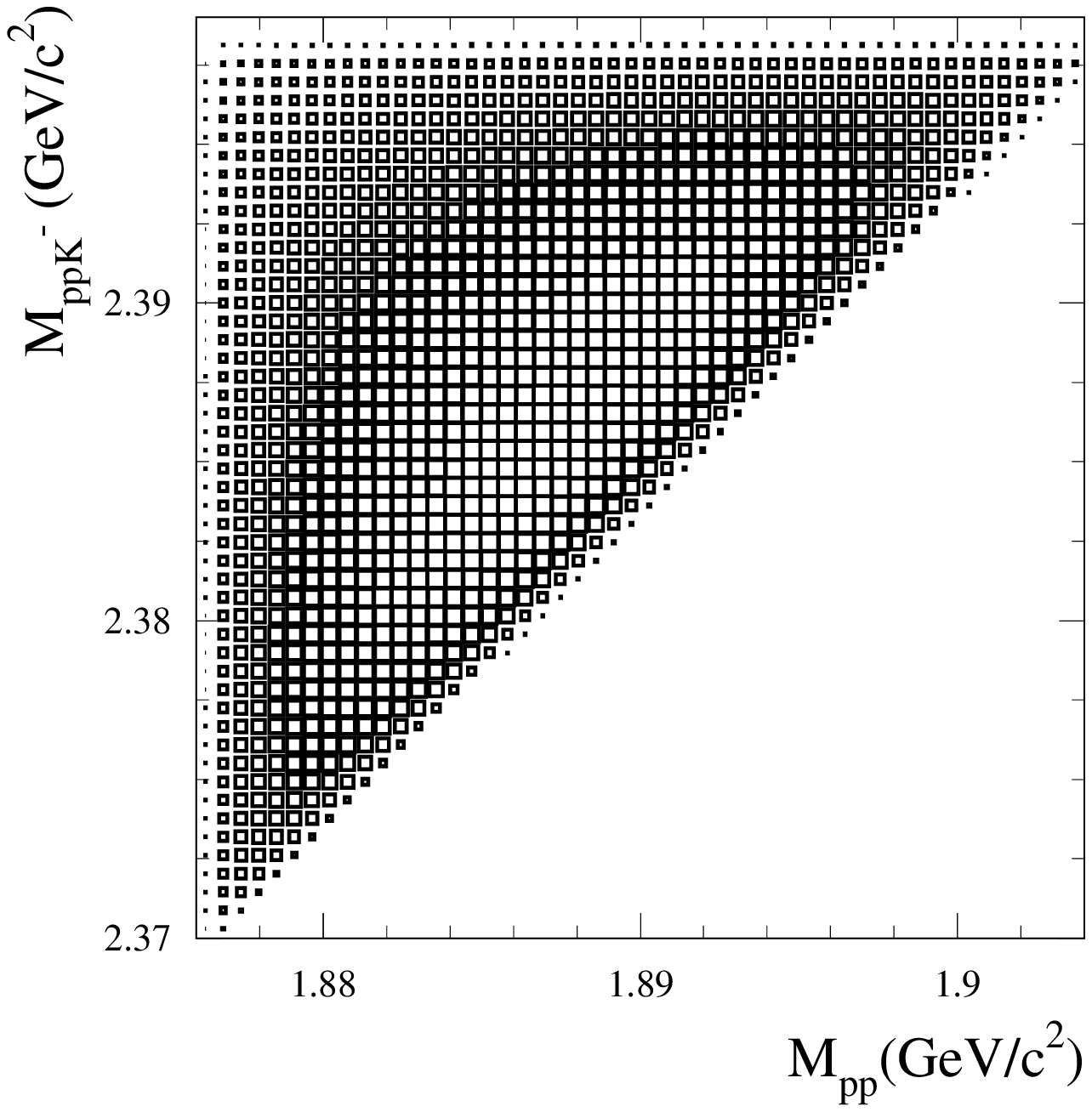,width=0.35\textwidth}}
\caption{Nyborg plot for the $pp\rightarrow ppK^{+}K^{-}$ reaction simulated at Q~=~28~MeV for 
the homogeneously populated phase space (left plot), for the case when proton-proton final
 state interaction is additionally taken into account (middle plot), and with only $pK^{-}$ final state interaction included (right plot).}
\label{mpp_mppkm_wyk}
\end{figure}  
From six two-particle and four three-particle invariant masses only five are independent 
due to the following relations\cite{nyborg}:
 \begin{equation}
\sum_{i,j=1 (i>j)}^{4}\,M_{ij}^{2} = s + 2 \sum_{i=1}^{4}\,m_{i}^{2}~,
\label{goldhaber1}
\end{equation}
\begin{equation}
M_{123}^{2} = M_{12}^{2} + M_{23}^{2} + M_{13}^{2} - m_{1}^{2} - m_{2}^{2} - m_{3}^{2}~,
\label{goldhaber2bis}
 \nonumber
\end{equation}
\begin{equation}
M_{134}^{2} = M_{13}^{2} + M_{34}^{2} + M_{14}^{2} - m_{1}^{2} - m_{3}^{2} - m_{4}^{2}~,
\label{goldhaber3bis}
\end{equation}
\begin{equation}
M_{124}^{2} = M_{12}^{2} + M_{24}^{2} + M_{14}^{2} - m_{1}^{2} - m_{2}^{2} - m_{4}^{2}~,
 \nonumber
\label{goldhaber4bis}
\end{equation}
\begin{equation}
M_{234}^{2} = M_{22}^{2} + M_{34}^{2} + M_{24}^{2} - m_{2}^{2} - m_{3}^{2} - m_{4}^{2}~.
 \nonumber
\label{goldhaber5bis}
\end{equation}
Under the assumption, that the matrix element depends only on invariant masses one can, as in the Chodrow approach, 
integrate equation~\eqref{prawdopodob} over the delta functions and spatial orientations. The result can be expressed
 as a distribution with some choice of the five independent invariant masses\cite{nyborg}. An especially convenient choice is
$M_{12}^{2},M_{34}^{2},M_{14}^{2},M_{124}^{2},$ and $M_{134}^{2}$\cite{nyborg}.
Using such variables, after integrations, one obtains an event distribution in the following form:
\begin{equation}
d^{5}P=\frac{\pi^{2}}{8s}\left|M\right|^{2}\frac{1}{\sqrt{-B}}~dM^{2}_{12}~dM^{2}_{34}~dM^{2}_{14}~dM^{2}_{124}~dM^{2}_{134}
\label{goldhaber3}
\end{equation}
Where $B$ is a function of the above-mentioned invariant masses, which exact form can be found in Nyborg's work~\cite{nyborg}.
An advantage of the choice of mass variables used here is the high symmetry of the function $B$\cite{nyborg}, which is very
usefull in consideration of the boundary of the physically allowed region in the $\left(M_{12}^{2},M_{34}^{2},M_{14}^{2},
M_{124}^{2},M_{134}^{2}\right)$ space defined by:
 \begin{equation}
B~=~0~.
\label{b}
\end{equation}
However, if the matrix element does not depend on all the selected invariant masses, distribution~\ref{goldhaber3} can be again 
integrated over this variables. Let suppose, that $M$ depends for example only on $M_{12}^{2},M_{34}^{2}$ and $M_{124}^{2}$,
which in case of the $pp\rightarrow ppK^{+}K^{-}$ reaction can be interpreted as $M_{pp}^{2},M_{K^{+}K^{-}}^{2}$ and $M_{ppK^{-}}^{2}$, 
respectively. This means that we assume that there are resonances only in subsystems characterized by the remaining invariant masses\cite{nyborg}. 
After integration over $M_{14}^{2}$ and $M_{134}^{2}$ one gets\cite{nyborg}:
\begin{equation}
d^{3}P=\frac{\pi^{3}}{8sM_{12}^{2}}\left|M\right|^{2}g\left(M_{12}^{2},m_{1}^{2},m_{2}^{2}\right)~dM^{2}_{12}~dM^{2}_{34}~dM^{2}_{124}
\label{goldhaber4}
\end{equation}
where $g$ is a function defined as:
\begin{equation}
g\left(x,y,z\right)=\sqrt{\left[x^{2}-\left(y-z\right)^{2}\right]\left[x^{2}-\left(y+z\right)^{2}\right]~.}
\label{goldhaber5}
\end{equation}
The physically allowed region is then a volume in the $\left(M_{12}^{2},M_{34}^{2},M_{124}^{2}\right)$ space 
bounded by a well defined, cusped surface\cite{nyborg}.\\
If we additionally change the integration variables to invariant masses the projection of 
the physical region on the $\left(M_{12},M_{34}\right)$-plane gives a convenient and simple 
distribution, which can be used to analyse the final state interaction in the same way as 
in case of three particles. Such analysis was originally made by Goldhaber et al.
in 1963\cite{goldhaber1,goldhaber2}. Therefore, after Nyborg et al.\cite{nyborg}, we will refer
to the above mentioned projection as to the Goldhaber plot.\\
As it is shown in Fig.~\ref{goldhaber_wyk} the physically allowed region on the Goldhaber plot is a right isosceles 
triangle within which the area is not proportional to the phase space volume\cite{nyborg}. 
Moreover the density of events on the Goldhaber plot in case of absence of any final state interaction
is also not homogeneous. However, it is still a very convenient tool due to its lorentz invariance and simple 
 boundary equations\cite{nyborg}:
 \begin{equation}
M_{12}+M_{34} = \sqrt{s}~,~M_{12} = m_{1}+m_{2}~,~M_{34} = m_{3}+m_{4}~.
\label{boundary1}
\end{equation}
It is worth mentioning that the projection on the $\left(M_{12},M_{124}\right)$-plane, hereafter 
referred to as the Nyborg plot\cite{nyborg,nyborg1},
 gives a very similar event distribution, which is again an isosceles triangle bounded by\cite{nyborg}:
\begin{equation}
M_{124} = M_{12}+m_{4}~,~M_{12} = m_{1}+m_{2}~,~M_{124} = \sqrt{s} - m_{3}~.
\label{boundary2}
\end{equation}
The other properties of this distribution are exactly the same as in the case of the 
Goldhaber plot (see Fig.~\ref{mpp_mppkm_wyk}).
\begin{figure}[H]
\centering 
\parbox{0.45\textwidth}{\centerline{\epsfig{file=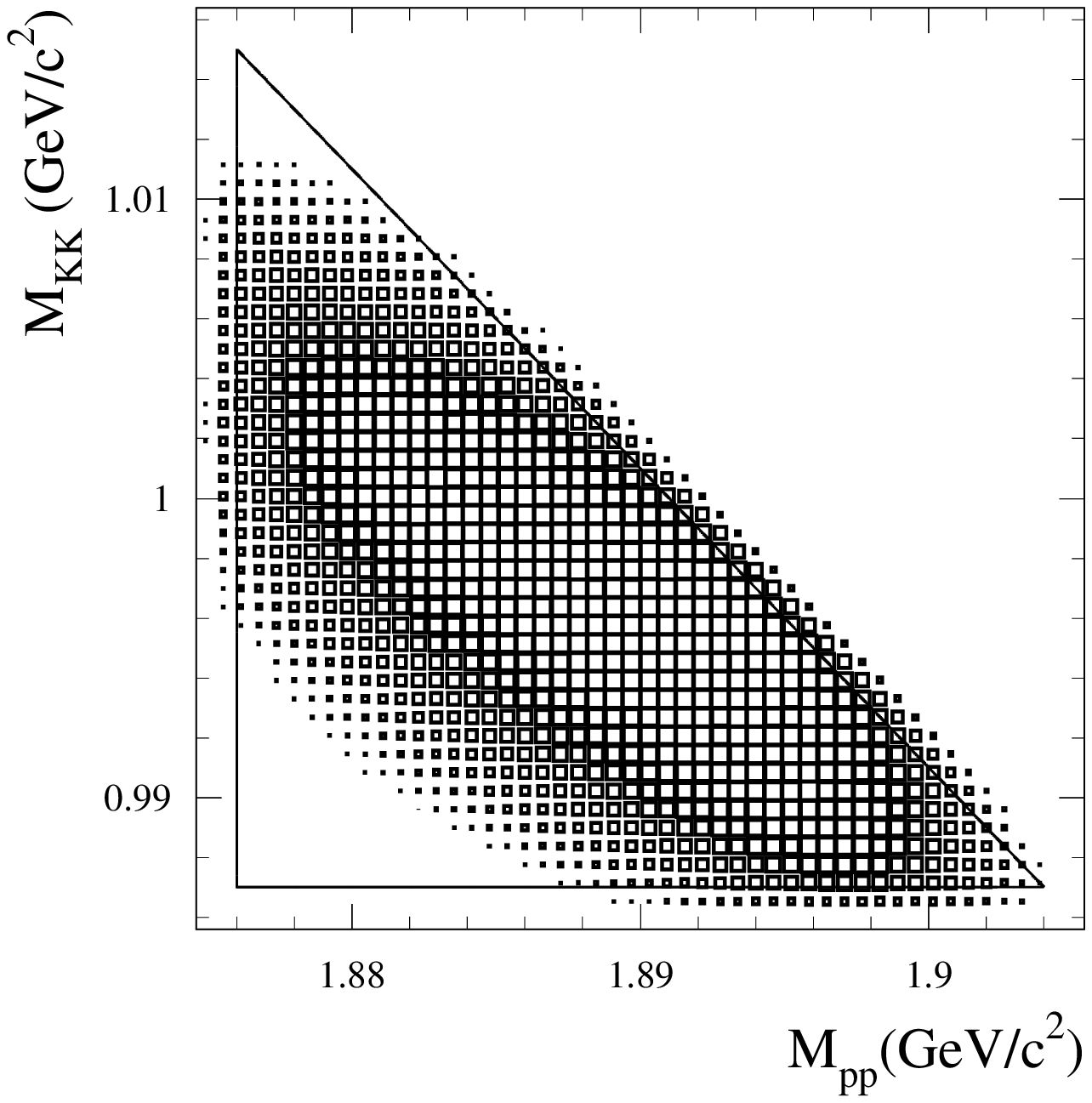,width=0.4\textwidth}}}
\parbox{0.45\textwidth}{\centerline{\epsfig{file=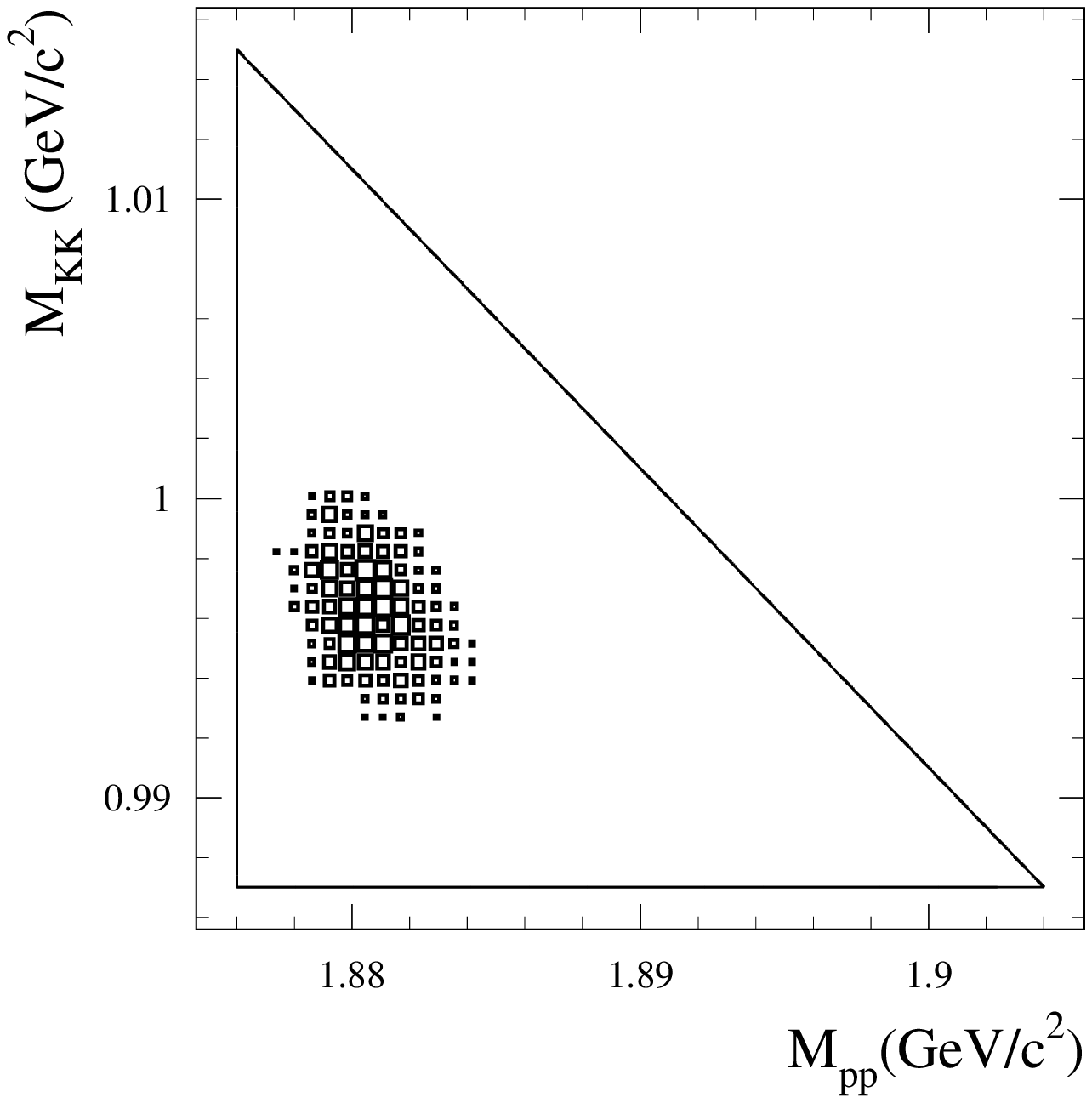,width=0.4\textwidth}}}
\parbox{0.45\textwidth}{\centerline{\epsfig{file=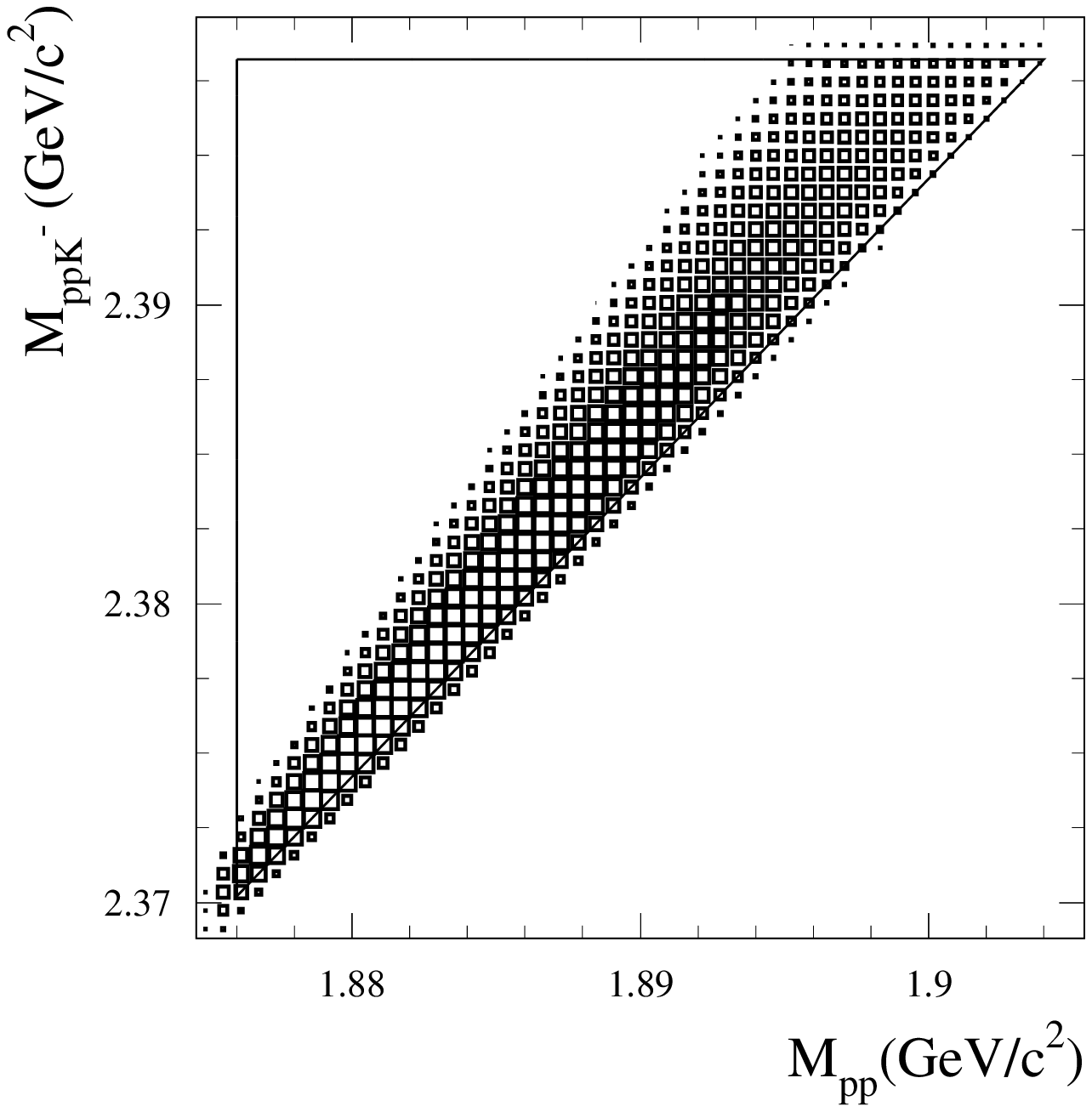,width=0.4\textwidth}}}
\parbox{0.45\textwidth}{\centerline{\epsfig{file=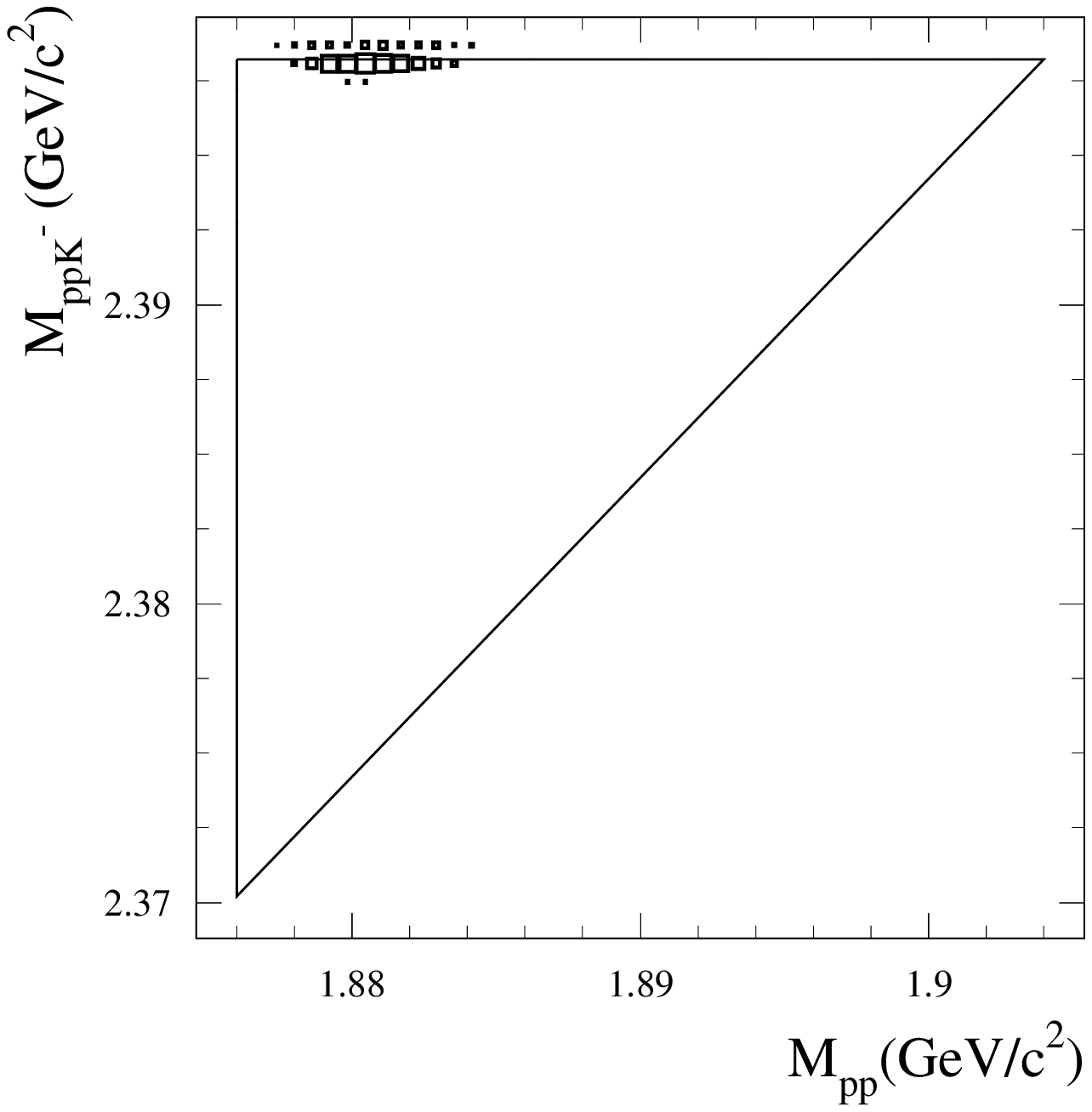,width=0.4\textwidth}}}
\caption{(\textbf{Upper panel}):~Goldhaber plot for the $pp\rightarrow ppK^{+}K^{-}$ 
reaction simulated at Q~=~28~MeV for a homogeneously populated phase 
space under the assumption that $M_{pK^{-}}$ is in the range from 1.432~MeV 
to 1.433~MeV (left plot) and in the range from 1.459~MeV to 1.460~MeV (right plot).~
(\textbf{Lower panel}): Nyborg plot for 
the $pp\rightarrow ppK^{+}K^{-}$ reaction simulated at Q~=~28~MeV for 
 a homogeneously populated phase space under the condition that 
$M_{pK^{-}}$ is in the range from 1.432~GeV/$c^{2}$ to 1.433~GeV/$c^{2}$ (left plot) and 
in the range from 1.459~GeV/$c^{2}$ to 1.460~GeV/$c^{2}$ (right plot).
\label{pasy}
}
\end{figure}  
Very interesting may be the comparison how the interaction between particles show up 
on Goldhaber and Nyborg plots. As it is depicted in Fig.~\ref{goldhaber_wyk} 
and Fig.~\ref{mpp_mppkm_wyk} (middle plots), for both plots the modification 
of event density due to $pp$--FSI appears to be very similar. On both distributions we observe a strong enhancement
in the region of small proton-proton invariant masses. However, when we take into account 
only the $pK^{-}$--FSI the situation changes 
and the modification of event density on the Goldhaber and Nyborg plots is small, but still visible. 
In order to study in more detail how the $pK^-$ interaction modifies the Goldhaber and Nyborg 
plots we simulated both distributions demanding, that the invariant mass of at 
least one $pK^{-}$ subsystem is in various ranges. In Fig.~\ref{pasy} 
we present results obtained under the assumption that the $M_{pK^{-}}$ 
is within a range from ($m_{p}~+~m_{K^{-}}$) to ($m_{p}~+~m_{K^{-}}$~+~0.001~GeV/$c^{2}$) 
and from ($m_{p}~+~m_{K^{-}}$~+~0.027~GeV/$c^{2}$) to ($m_{p}~+~m_{K^{-}}$~+~0.028~GeV/$c^{2}$).
It is clearly seen, that under such assumptions the physically allowed region on the Goldhaber plot is 
much bigger than on the Nyborg plot, especially for low $M_{pK^{-}}$ invariant masses. This means, that in the 
region, where $pK^{-}$--FSI is expected to be strong, events are confined to a small part of the Nyborg plot 
and are distributed on much larger surface on the Goldhaber plot. That is why one expects larger effects from the 
$pK^-$ final state interaction in the Nyborg plot then in the Goldhaber distribution.
 It is worth mentioning, that the above described differences between this two 
generalizations of the Dalitz plot suggest, that the Goldhaber plot is more appropriate in the analysis of 
the $K^{+}K^{-}$ interaction, while in the investigation of the $pK^{-}$--FSI one should use the Nyborg distribution.\\ 
Using the symmetry of the $B$ function in equation \eqref{goldhaber3}, several other two-dimensional distributions 
can be obtained\cite{nyborg}. However in the analysis of the $K^{+}K^{-}$ interaction, which is presented in the 
next chapter, we use only the two distributions described in this section. 
\pagestyle{myheadings}
\chapter{Study of the $K^{+}K^{-}$ final state interaction}
\hspace{\parindent}
\section{Experimental Goldhaber and Nyborg plots}
\hspace{\parindent}
As it was mentioned in the previous chapters, the analysis of the $K^{+}K^{-}$ final state interaction is based on 
data obtained in COSY-11 measurements at two beam momenta. For both energies 
we chose events identified as a $pp\rightarrow ppK^{+}X^{-}$ reaction 
with missing mass in the region of the kaon (0.235~Ge$V^{2}/c^{4}$~<~$m^{2}_{X}$~<~0.25~Ge$V^{2}/c^{4}$)\cite{cosy2}.
Knowing the four-momenta of protons and kaons for each event one can calculate appropriate invariant 
masses and construct the experimental event distributions as described in the previous chapter.
\begin{figure}[H]
\centering 
\parbox{0.45\textwidth}{\centerline{\epsfig{file=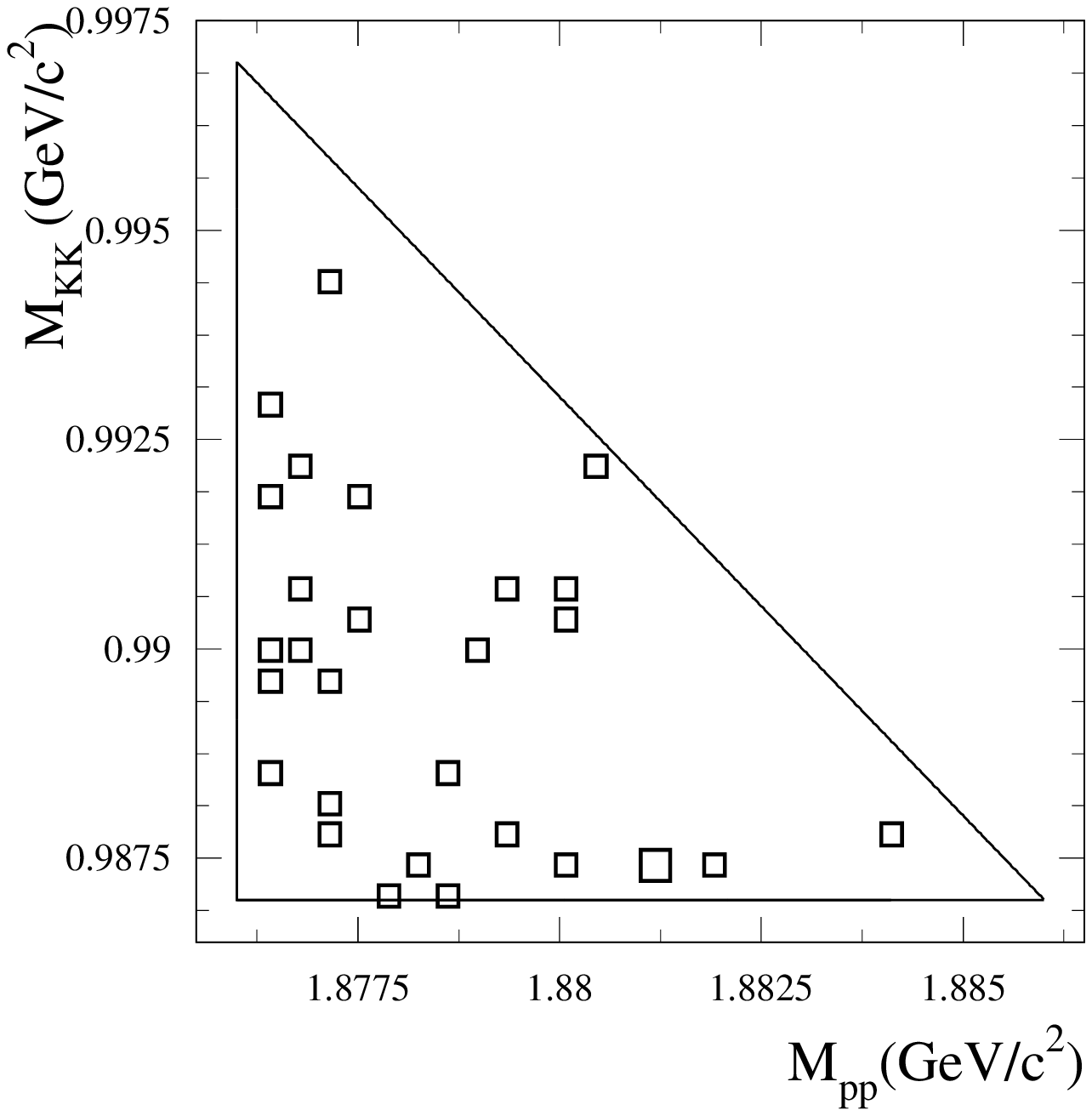,width=0.4\textwidth}}}
\parbox{0.45\textwidth}{\centerline{\epsfig{file=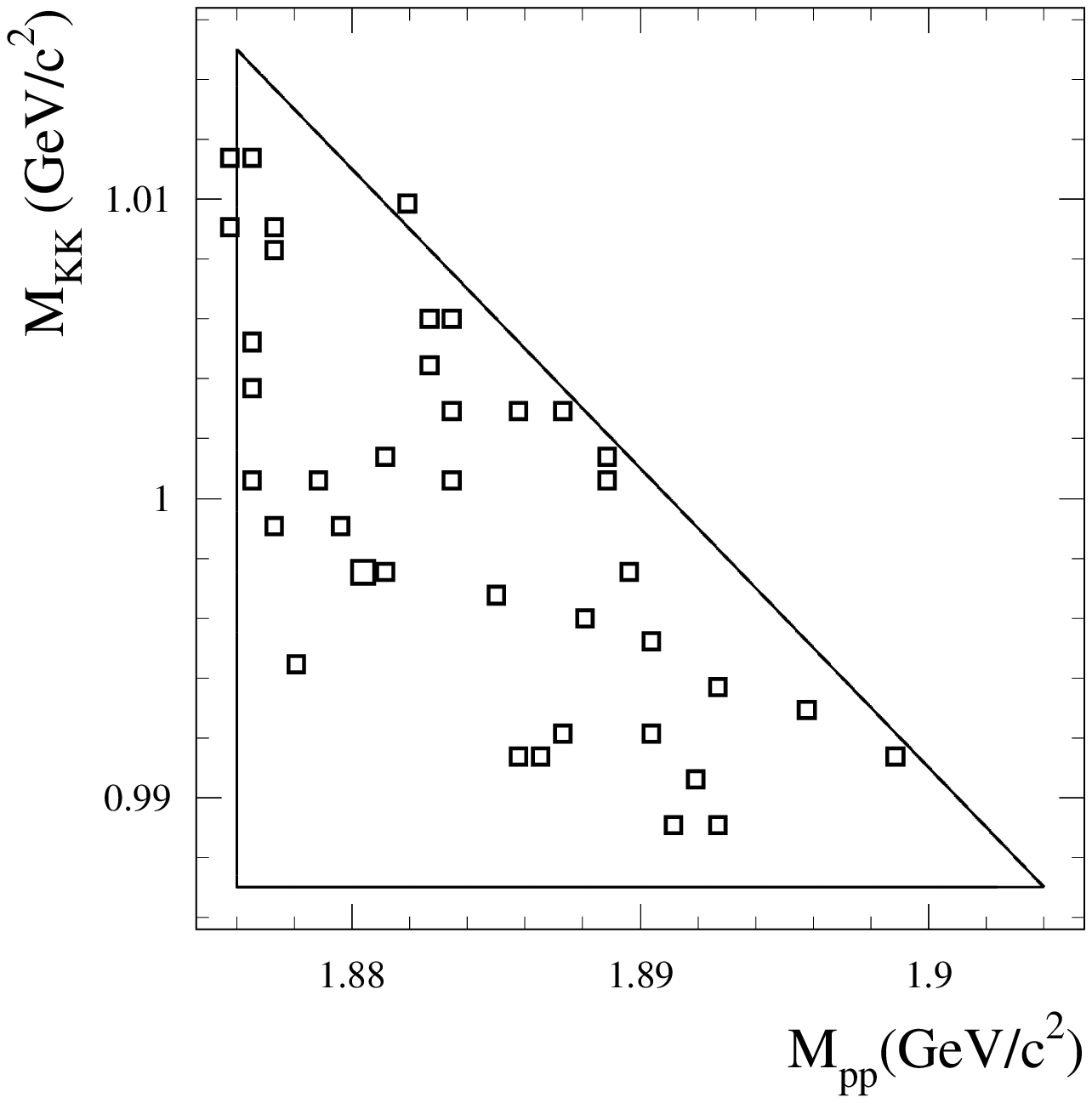,width=0.4\textwidth}}}
\parbox{0.45\textwidth}{\centerline{\epsfig{file=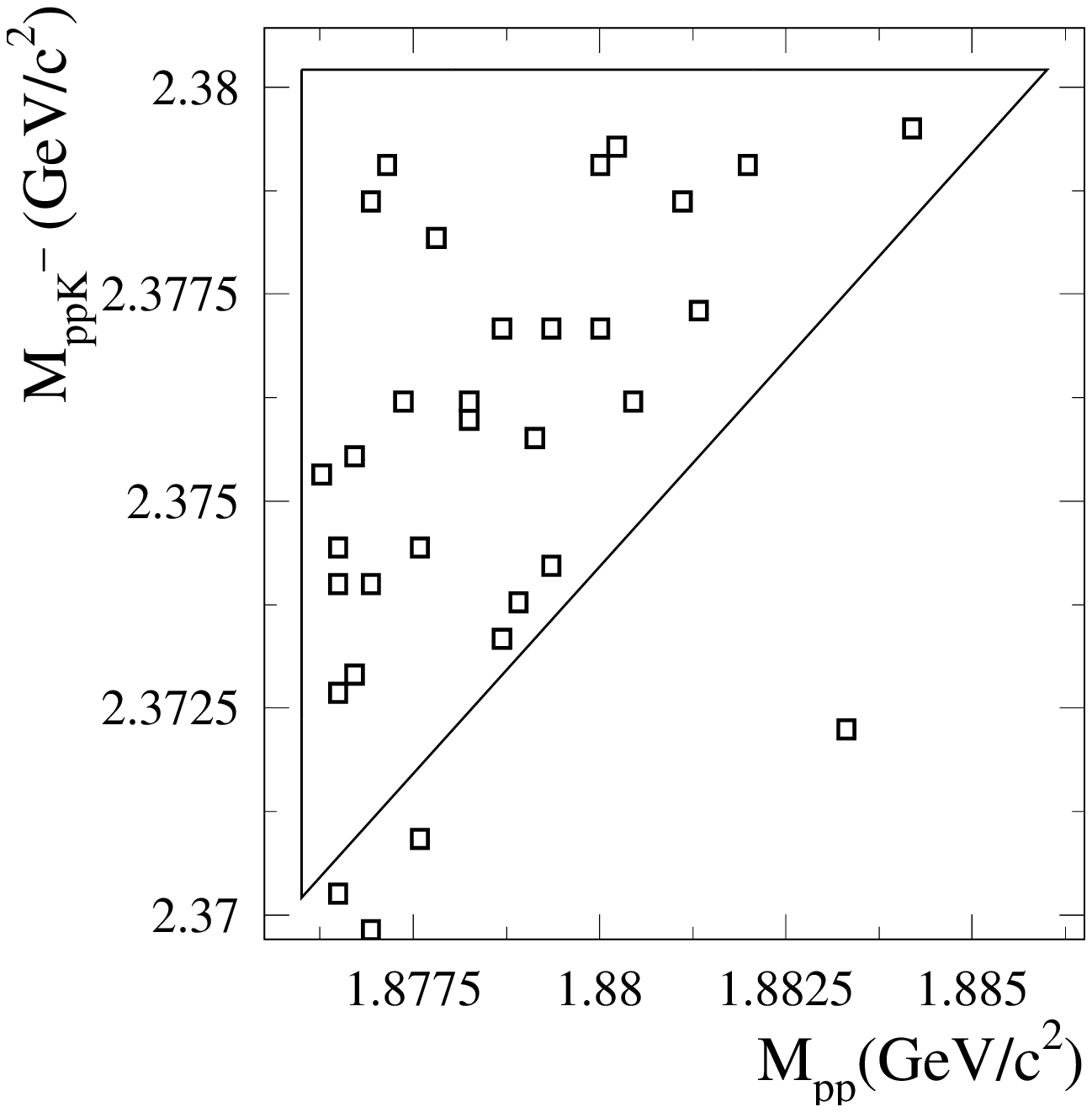,width=0.4\textwidth}}}
\parbox{0.45\textwidth}{\centerline{\epsfig{file=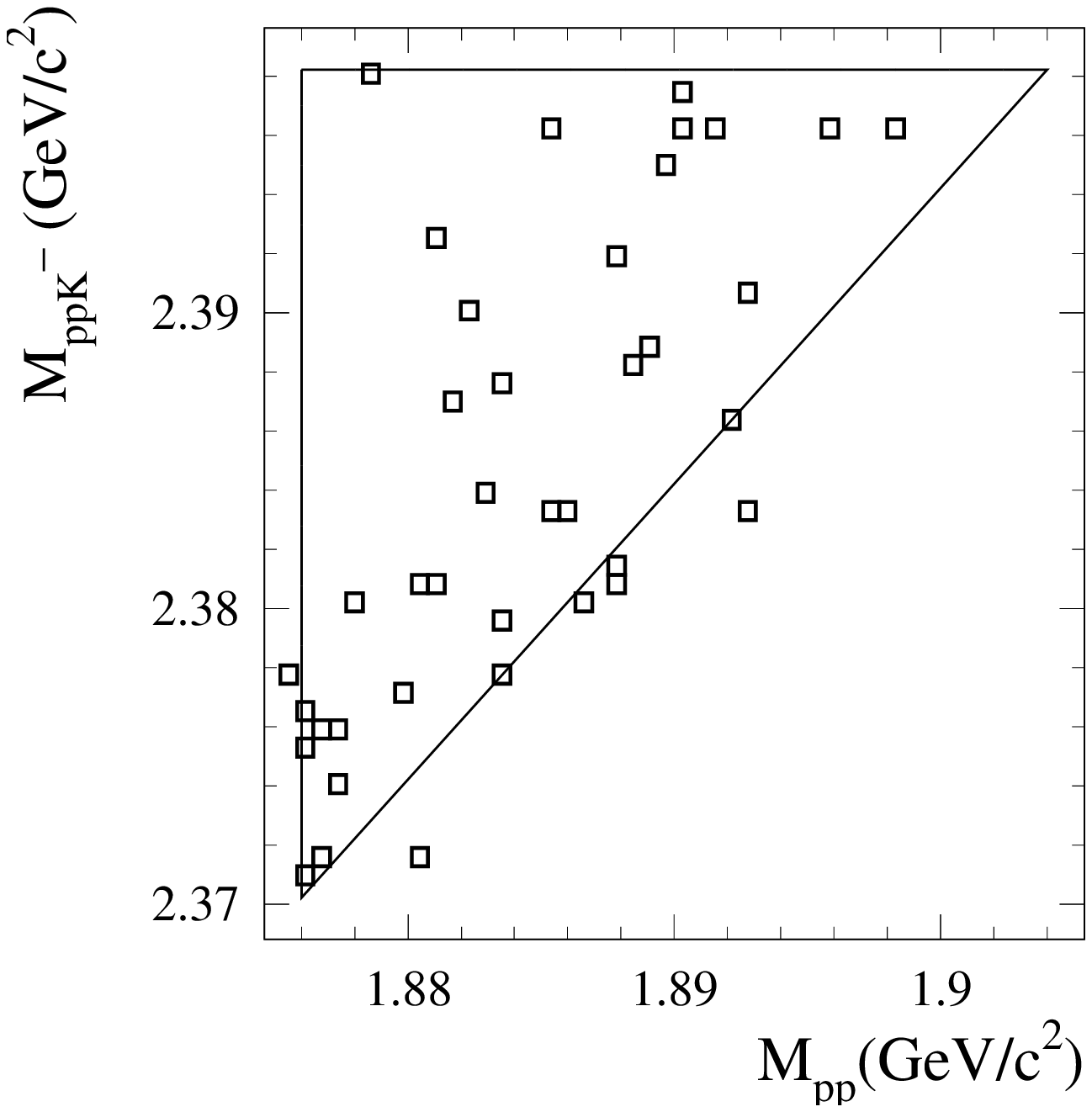,width=0.4\textwidth}}}
\caption{(\textbf{Upper panel}):~Goldhaber plots obtained from measurements 
at Q~=~10~MeV (left) and Q~=~28~MeV (right). (\textbf{Lower panel}):~Nyborg plots obtained 
from measurements at Q~=~10~MeV (left) and Q~=~28~MeV (right).
Solid lines show boundaries of the physically allowed region.}
\label{goldhaber_exp}
\end{figure}  
Fig.~\ref{goldhaber_exp} shows raw experimental event distributions determined at both excess energies. 
These distributions need to be corrected for the acceptance and detection 
efficiency of the COSY-11 facility. Only then it will be possible to extract the experimental 
differential cross sections. 
The derived cross sections will be then compared to the results of the Monte Carlo simulations 
generated with various parameters of $K^{+}K^{-}$ interaction under the 
assumption that there is an additional strong final state interaction in $pp$ and $pK^{-}$ subsystems\cite{c_wilkin}.
A more detailed description is presented in further parts of this chapter.
\section{Corrections for the acceptance and detection efficiency}
\begin{figure}[H]
\centering 
\parbox{0.4\textwidth}{
{\epsfig{file=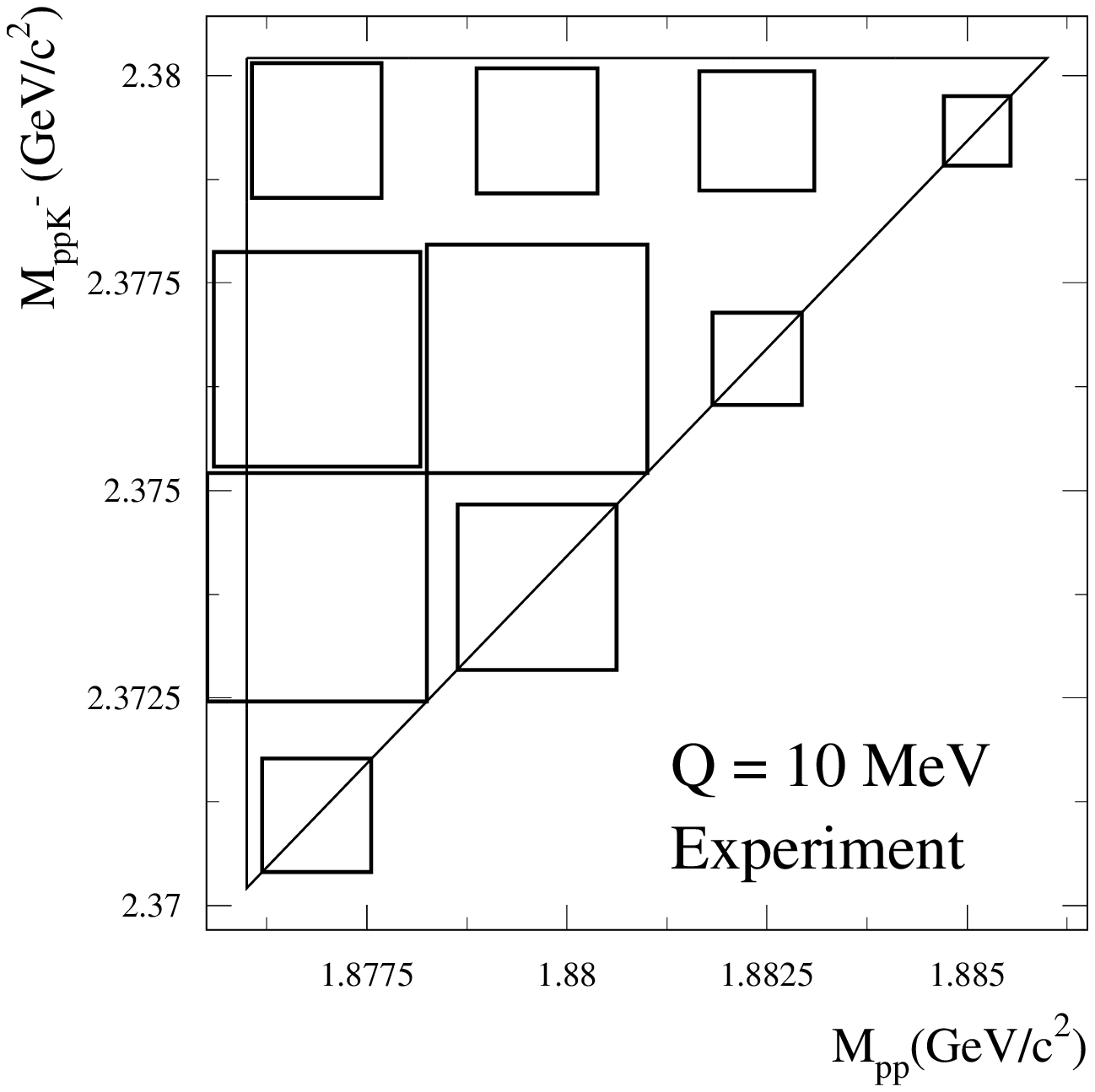,width=0.4\textwidth}}}
\parbox{0.4\textwidth}{
{\epsfig{file=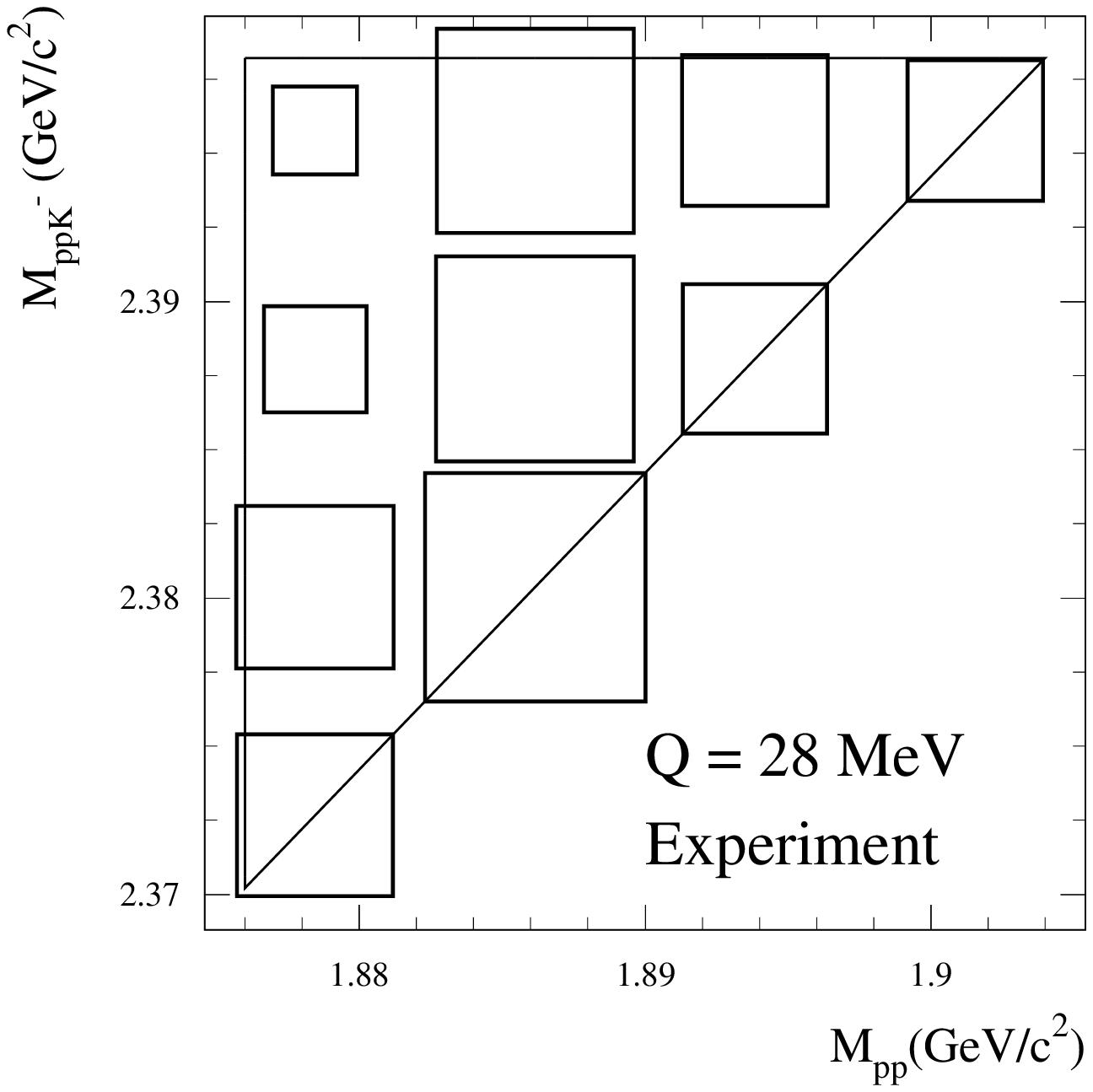,width=0.4\textwidth}}}
\parbox{0.4\textwidth}{
{\epsfig{file=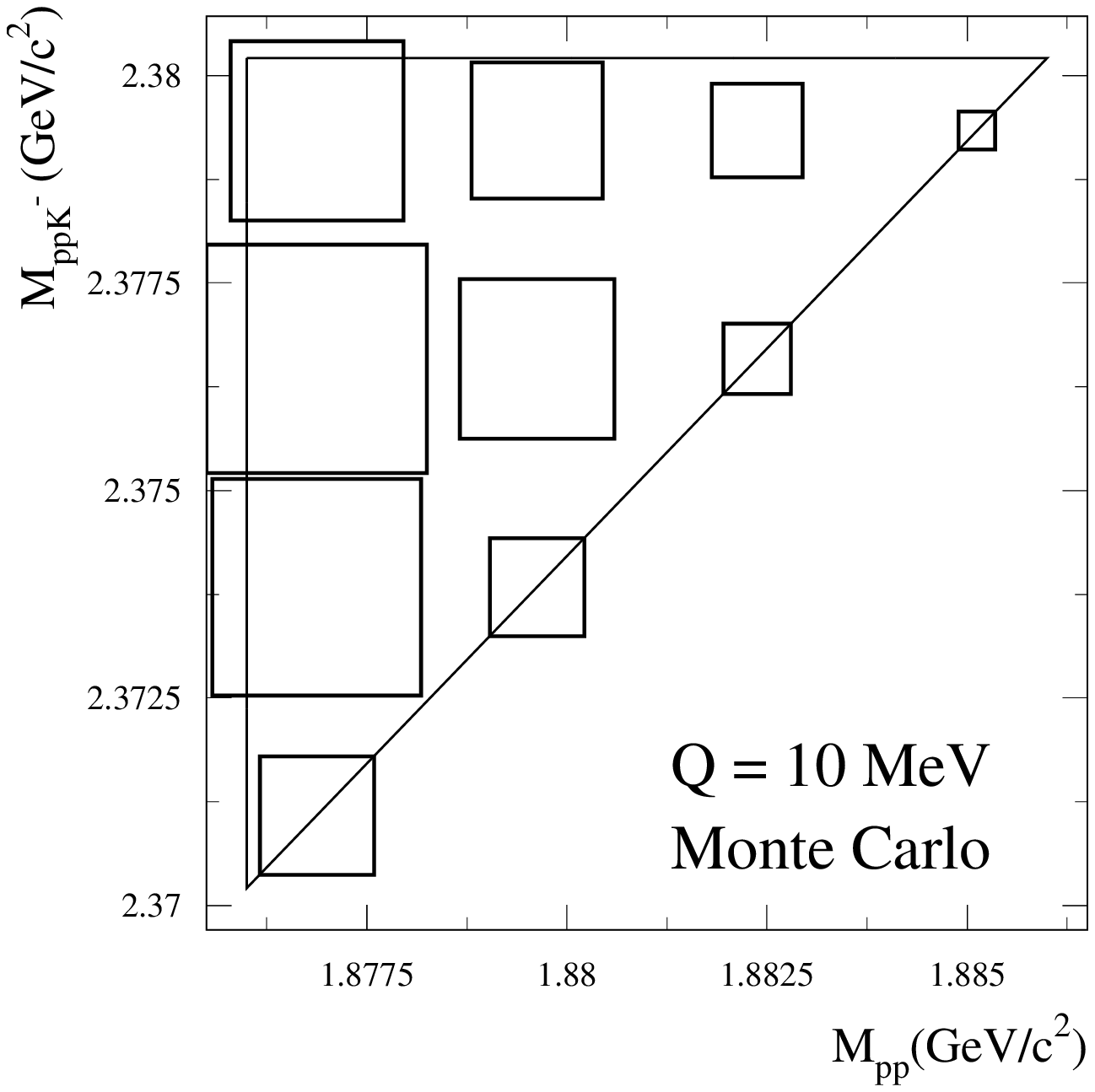 ,width=0.4\textwidth}}}
\parbox{0.4\textwidth}{
{\epsfig{file=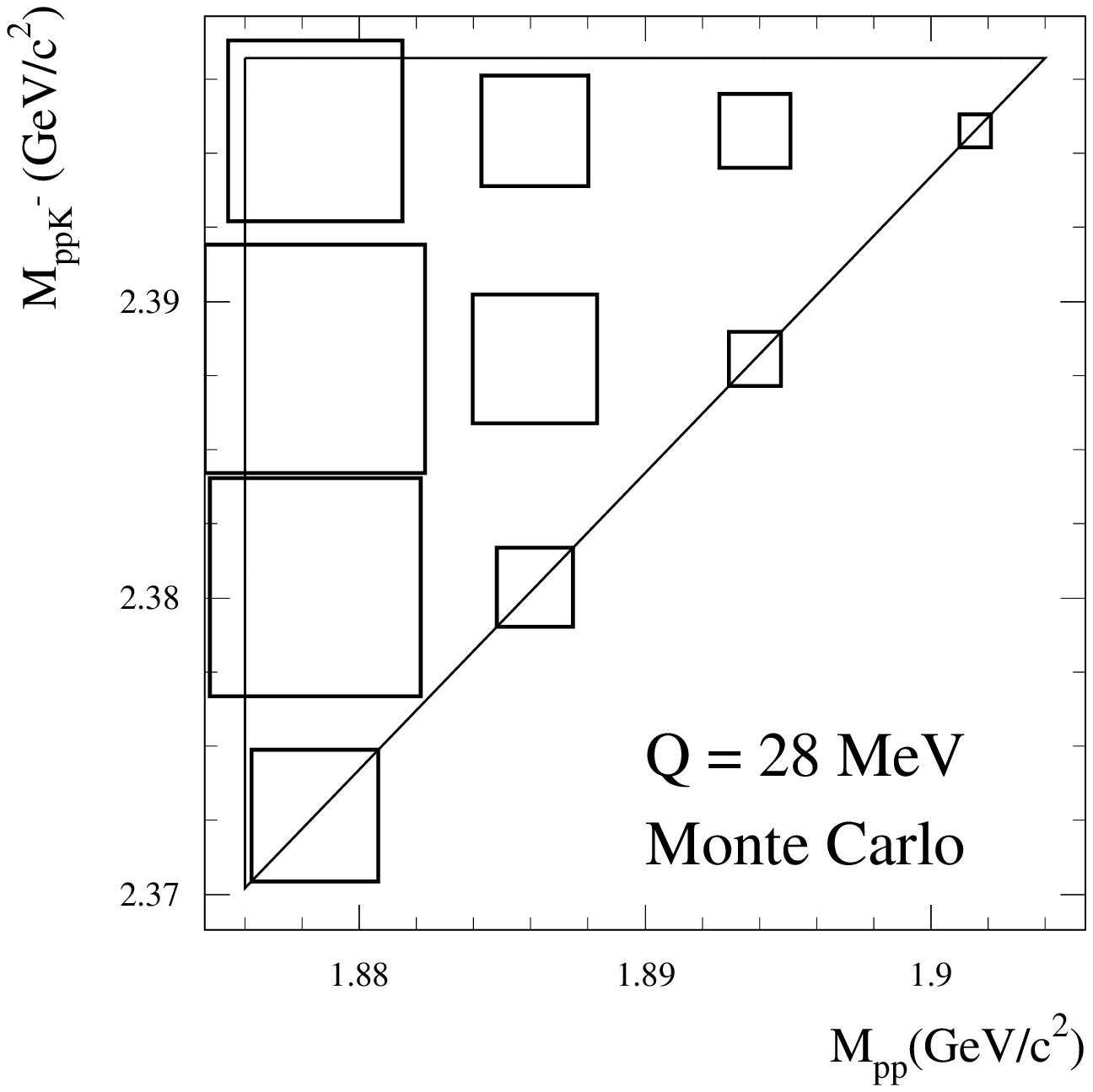 ,width=0.4\textwidth}}}
\caption{(\textbf{Upper panel}):~Experimental Nyborg plots after corrections for the COSY-11 
acceptance for the $pp\rightarrow ppK^+K^-$
reaction determined at Q~=~10~MeV (left) and Q~=~28~MeV (right). 
(\textbf{Lower panel}): Event distributions obtained from simulations of the $pp\rightarrow
ppK^{+}K^{-}$ reaction at Q~=~10~MeV (left) and Q~=~28~MeV (right).
\label{goldhaber_przekroje1}
}
\end{figure}

\hspace{\parindent}
In order to obtain the experimental double differential cross sections~$\frac{d^2\sigma\left(M_{pp},M_{K^{+}K^-}\right)}
{dM_{pp}~dM_{K^{+}K^-}}$, the data were first binned 
into intervals of $\Delta$M~=~2.5~MeV width for the measurement at Q~=~10~MeV and intervals of
$\Delta$M~=~7~MeV for the data at Q~=~28~MeV. 
The acceptance corrections were made separately for each bin according to the formulae which
will be derived in this section.\\  
The number of experimental events in a given bin 
can be expressed as:
\begin{equation}
\frac{\Delta N_{exp}\left(M_{pp},M_{K^+K^-}\right)}{\Delta M_{pp}~\Delta M_{K^+K^-}}~=
~\frac{d^2\sigma\left(M_{pp},M_{K^+K^-}\right)}{dM_{pp}~dM_{K^+K^-}}~L_{int}~A\left(M_{pp},M_{K^+K^-}\right)~,
\label{acc1}
\end{equation}
where, $A\left(M_{pp},M_{K^{+}K^-}\right)$ denotes the COSY-11 acceptance and detection efficiency for the 
coincident measurement of protons and kaons, 
with invariant masses in the range: 
\begin{equation}
M_{pp}\in \left(M_{pp}~-~\frac{\Delta M_{pp}}{2};~M_{pp}~+~\frac{\Delta M_{pp}}{2}\right)
 \nonumber
\end{equation}
\begin{equation} 
M_{KK}\in \left(M_{KK}~-~\frac{\Delta M_{KK}}{2};~M_{KK}~+~\frac{\Delta M_{KK}}{2}\right)~,
 \nonumber
\end{equation}
and where $L_{int}$ denotes the luminosity integrated over whole time of the measurement:
\begin{equation}
L_{int}~=~\int L(t)dt~. 
\label{l}
\end{equation}
From eq.~\eqref{acc1} one can calculate the double differential cross section:
 \begin{equation}
\frac{d^2\sigma\left(M_{pp},M_{K^{+}K^{-}}\right)}{dM_{pp}~dM_{K^+K^-}}~=~\frac{\Delta N_{exp}\left(M_{pp},M_{K^+K^-}\right)}
{\Delta M_{pp}~\Delta M_{K^+K^-}}\frac{1}{L_{int}~A\left(M_{pp},M_{K^{+}K^{-}}\right)}~.
\label{acc2}
\end{equation}
The values of the integrated luminosity amount to $L_{int}$~=~2.770~$\pm$~0.056~pb$^{-1}$ for the measurement at 
Q~=~10~MeV and $L_{int}$~=~2.270~$\pm$~0.007~pb$^{-1}$ at Q~=~28~MeV\cite{cosy2}. To determine the acceptance we first generated 
five milions of $pp\rightarrow ppK^{+}K^{-}$ events using a FORTRAN-based code, called GENBOD\cite{genbod}. 
It permits to generate four-momentum vectors of
the outgoing particles in the centre of mass frame with a homogeneous distribution in the phase space. 
The total centre of mass energy as well as the number and masses of the particles are specified by the 
user\cite{silar}. 
Next for each generated event a response of the COSY-11 detection system was calculated using 
a GEANT 3 based software package\cite{geant}.
\begin{figure}[H]
\centering 
\parbox{0.4\textwidth}{\centerline{\epsfig{file=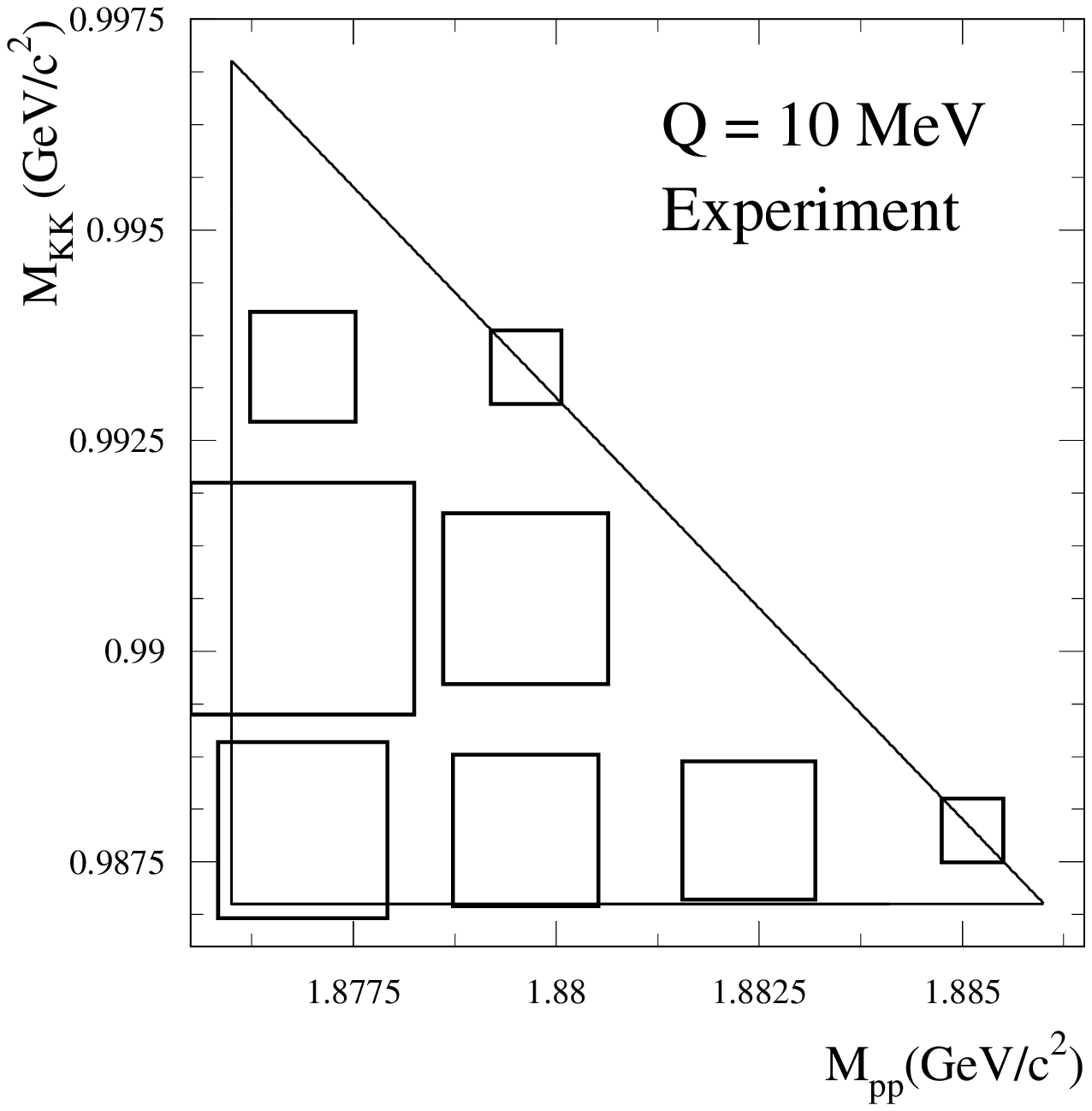,width=0.4\textwidth}}}
\parbox{0.4\textwidth}{\centerline{\epsfig{file=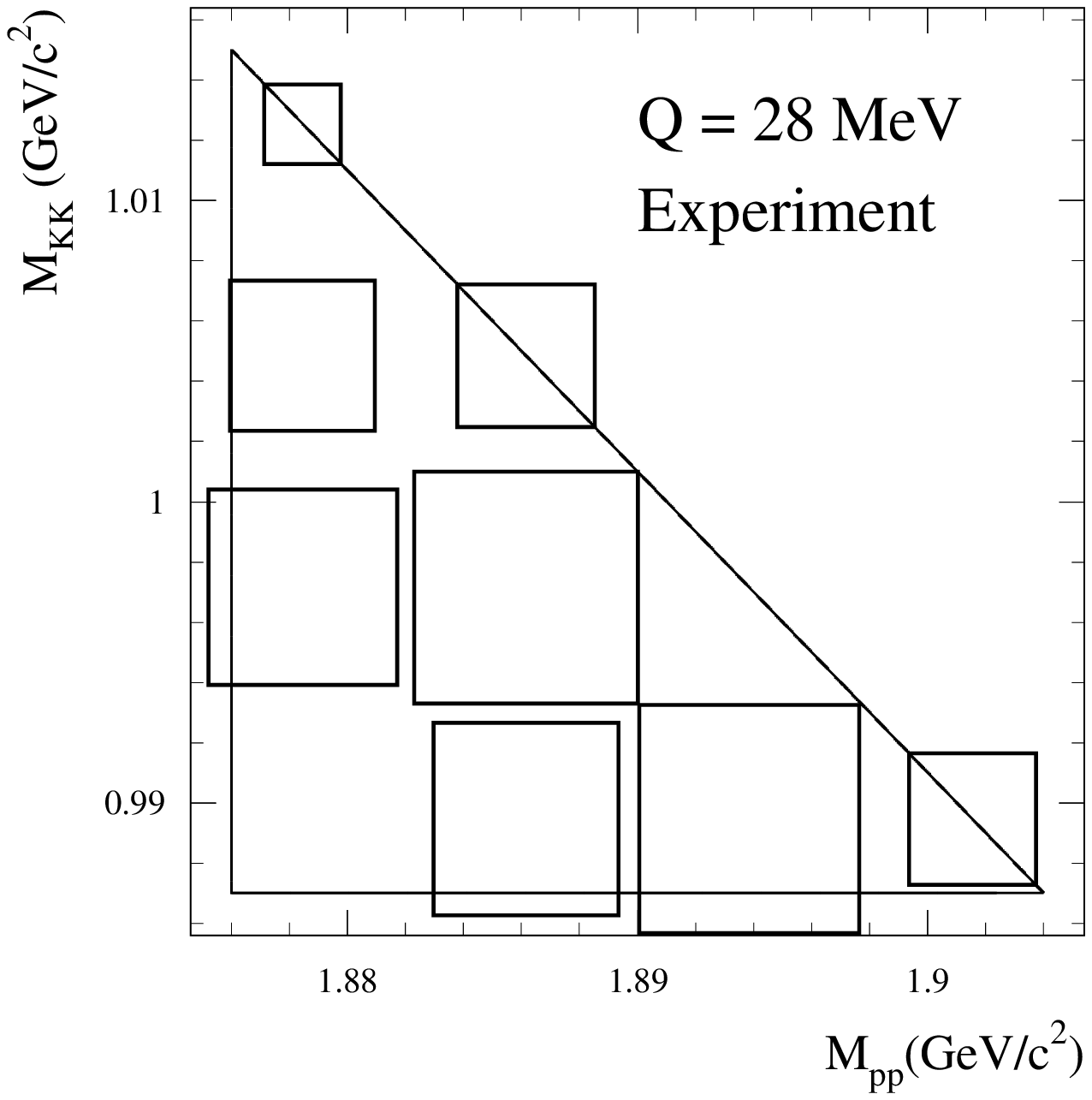,width=0.4\textwidth}}}
\parbox{0.4\textwidth}{\centerline{\epsfig{file=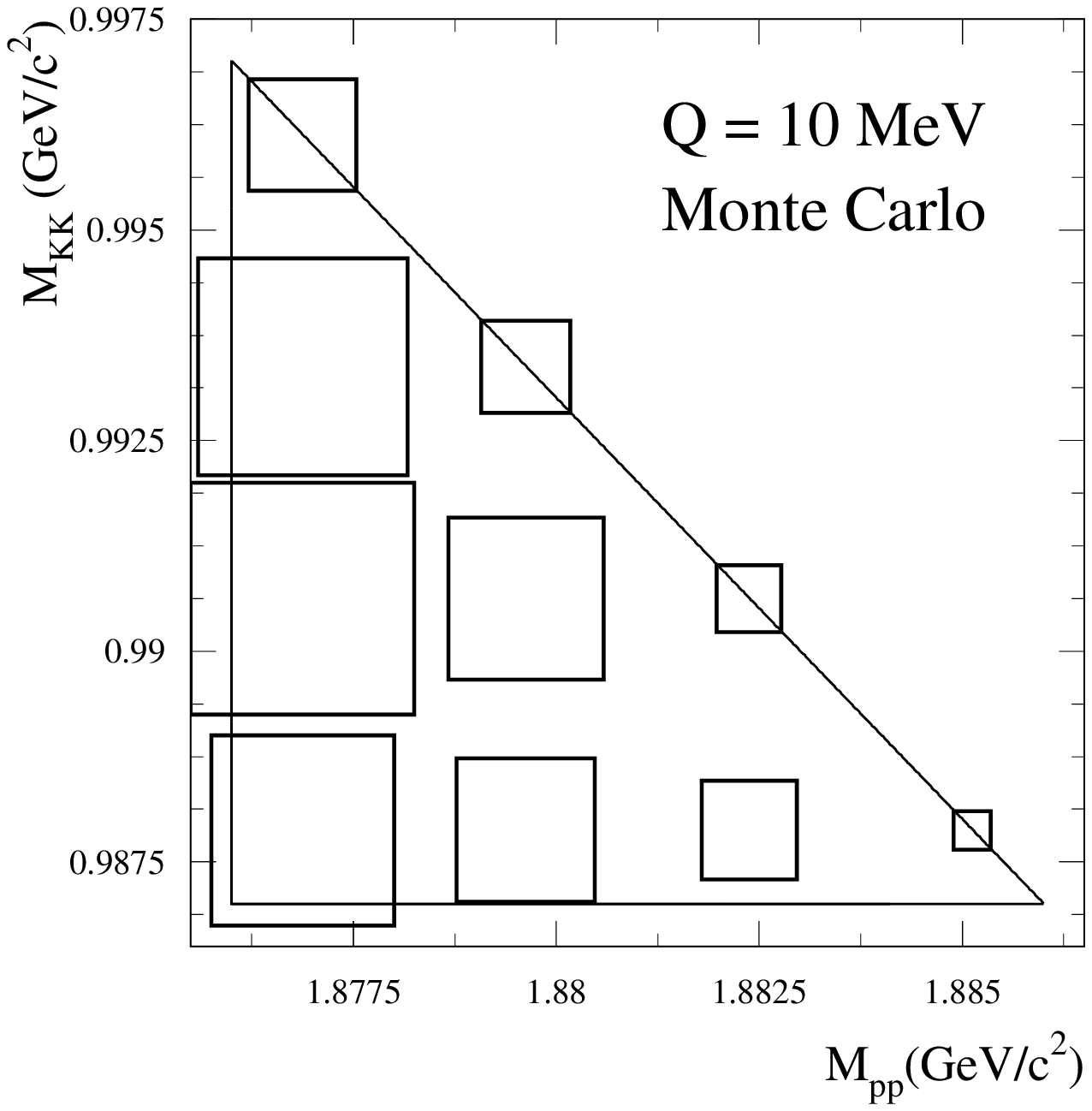,width=0.4\textwidth}}}
\parbox{0.4\textwidth}{\centerline{\epsfig{file=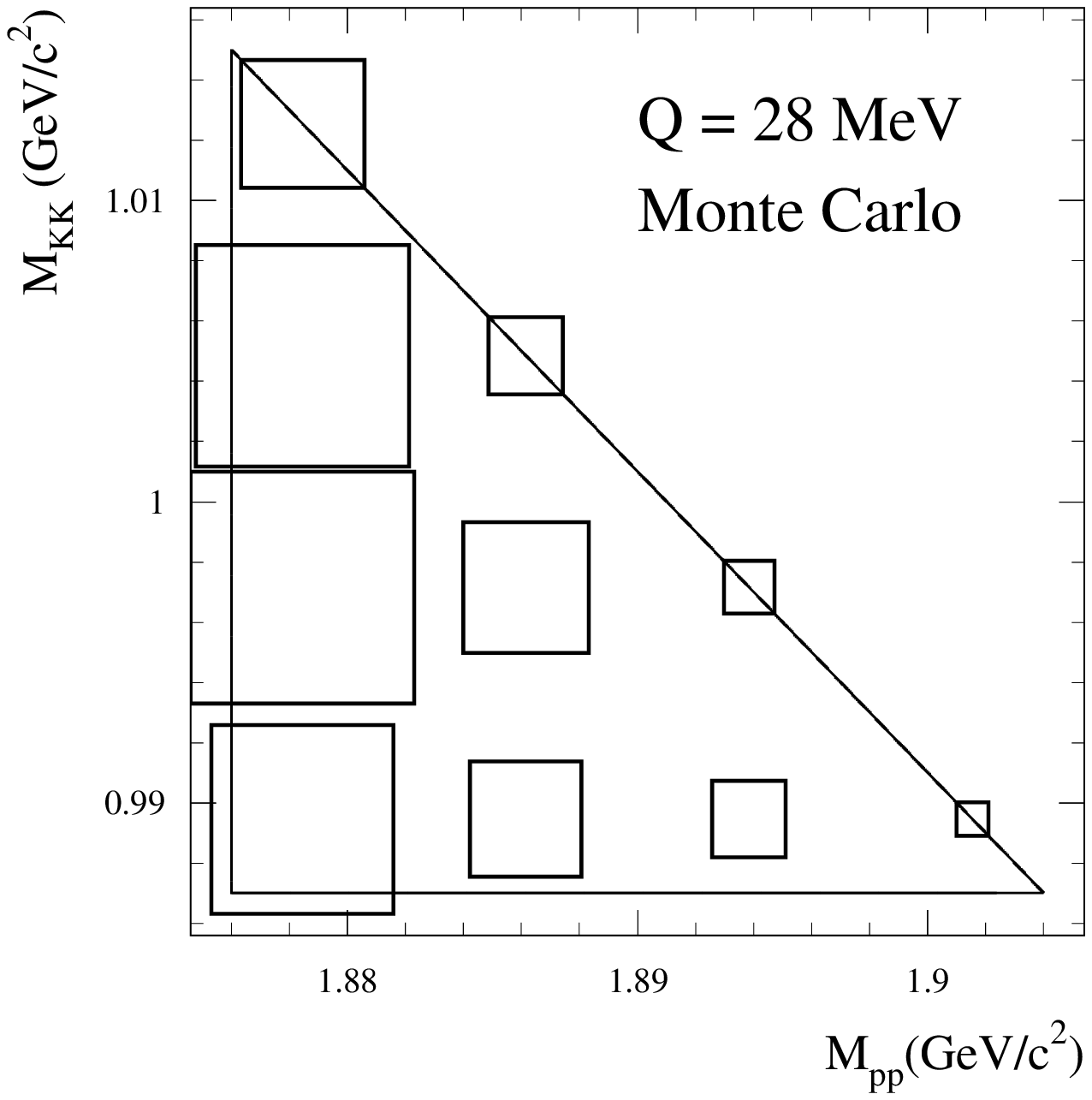,width=0.4\textwidth}}}
\caption{(\textbf{Upper panel}):~Experimental Goldhaber plots obtained after corrections for acceptance and detection 
efficiency from measurements 
at Q~=~10~MeV (left) and Q~=~28~MeV (right). (\textbf{Lower panel}): Goldhaber plots simulated 
at Q~=~10~MeV (left) and Q~=~28~MeV (right) with $pp$--FSI included.
\label{goldhaber_przekroje}
}
\end{figure}
Then we reconstructed the momenta and energies of the particles applying the same program 
which had been used for analysis of the experimental data. The generated and reconstructed events were binned 
exactly in the same way as it was described for the experimental data.\\
In order to account for the $pp$--FSI each event was weighted by the square of the 
proton-proton scattering amplitude expressed as\cite{habilitacja,pp-FSI}:
\begin{equation}
\label{amppp}
F_{pp} = 
  \frac{e^{-i\delta_{pp}({^{1}\mbox{\scriptsize S}_{0}})} \cdot 
        \sin{\delta_{pp}({^{1}\mbox{S}_0})}}
       {C \cdot \mbox{k}}~,
\end{equation}
where $C$ stands for the square root of the Coulomb penetration factor\cite{pp-FSI}, 
and k represents  either of the proton momentum in the proton-proton rest frame. 
The parameter $\delta_{pp}({^{1}\mbox{S}_0})$ denotes the phase-shift calculated according
 to the modified Cini-Fubini-Stanghellini formula with the Wong-Noyes Coulomb 
correction\cite{noyes995,naisse506,noyes465}. A more detailed description of this $pp$--FSI
parametrization can be found in references\cite{habilitacja,pp-FSI,noyes995,naisse506,noyes465}.\\
Finally the acceptance $A\left(M_{pp},M_{K^{+}K^{-}}\right)$ was calculated using 
following formula:
\begin{equation}
A\left(M_{pp},M_{K^{+}K^{-}}\right)~=~\frac{\displaystyle \sum_{i} w_{i}^{rec}\left(M_{pp},M_{K^{+}K^{-}}\right)}
{\displaystyle \sum_{j}w_{j}^{gen}\left(M_{pp},M_{K^{+}K^{-}}\right)}~,
\label{acc4}
\end{equation}
where $w_{i}^{rec}\left(M_{pp},M_{K^{+}K^{-}}\right)$ denotes the weight for the $i$-th 
reconstructed event in a specific $\left(M_{pp},M_{K^{+}K^{-}}\right)$ bin. 
Analogously $w_{j}^{gen}\left(M_{pp},M_{K^{+}K^{-}}\right)$ stands for weight 
for the $j$-th generated event in the same bin.\\
Knowing the acceptance of the COSY-11 detection setup and using eq.~\eqref{acc2} we obtained 
distributions of the differential cross section. The results are presented
in Fig.~\ref{goldhaber_przekroje} and \ref{goldhaber_przekroje1}.
A qualitative comparison between measured and expected distributions indicates, that inclusion of
the $pp$--FSI is not sufficient to fully describe the experimental data. 
If the $pK^{-}$ and $K^+K^-$ were negligible, in the experimental distributions one would observe a big enhancement
only in the region of small $M_{pp}$, where protons have small relative momenta as it is seen in the
simulations. But for the measurements we obtained a significant increase of the event density also in the region of small $M_{KK}$ 
(Fig.~\ref{goldhaber_przekroje}) and big $M_{ppK^-}$ (see Fig.~\ref{goldhaber_przekroje1}), which
may be a manifestation of the $K^{+}K^{-}$ or $pK^{-}$ interaction. Moreover in both experimental
 distributions 
 the enhancement expected from $pp$--FSI is shifted towards higher proton-proton invariant
masses.\\ 
However, in order to draw conclusions in the next section we will carry out the quantitative 
comparison between analysed data and theoretical distributions simulated with different parameters describing
the final state interaction. 
\section{Determination of the scattering length of the $K^{+}K^{-}$ interaction}
\hspace{\parindent}
To describe the experimental data in terms of final state interactions between 
 the two protons, $K^-$ and  protons and kaon and antikaon, we made 
 the assumption that the overall final state interaction enhancement factor can 
 be expressed as a following product:
 \begin{equation}
\label{fsi-factor}
F_{FSI}~=~F_{pp}~\times~F_{ppK^{-}}~\times~F_{K^+K^-}~,
\end{equation}
where $F_{ppK^-}$ and $F_{K^+K^-}$ denote the enhancement factor in the
$ppK^-$ and $K^+K^-$ subsystems respectively. These factors can be expressed 
in the scattering length approximation as\cite{c_wilkin,anke}:
\begin{equation}
  \nonumber
\label{scattering_length}
F_{ppK^{-}}~=~\frac{1}{(1-ik_{1}a_{pK^{-}})(1-ik_{2}a_{pK^{-}})}~,
\end{equation}
\begin{equation}
\label{scattering_length1}
F_{K^+K^-}~=~\frac{1}{(1-iqa_{K^{+}K^{-}})}~,
\end{equation}
where $k_1$, $k_2$ and $q$ stands for relative momenta of particles in the first $pK^-$ subsystem,
second $pK^-$ subsystem and $K^+K^-$ subsystem respectively. $a_{pK^{-}}$ and $a_{K^{+}K^{-}}$ 
stands for the effective scattering length of the appropriate interacting pair.
In investigations described in this thesis we assumed that the $pK^-$ scattering length 
amounts to: $a_{pK^{-}}$~=~(0~+~1.5$i$) fm as it was established by the ANKE group\cite{anke}.\\
Using the mentioned parametrizations of the final state interactions we 
compared the experimental event distributions to the results of Monte Carlo
simulations treating the $K^+K^-$ scattering length as an unknown parameter, which 
has to be determined. For finding an estimate of the real and imaginary part of 
$a_{K^{+}K^{-}}$ we constructed the $\chi^{2}$ statistic according to the {\em method of least
squares}:
\begin{equation}
\chi^2(\alpha,a_{K^{+}K^{-}}) =  \sum_i \, \frac{\left(\frac{d^2\sigma_{i}}{dM_{pp}~dM_{K^+K^-}} 
- \alpha N_i^{MC}(a_{K^{+}K^{-}})\right)^{2} }{\delta_{i}^{2}}~,
\label{eqchi2_sl}
\end{equation}
where $\frac{d^2\sigma_{i}}{dM_{pp}~dM_{K^+K^-}}$ denotes the differential cross section determined 
for a $i^{th}$ bin of the experimental Goldhaber or Nyborg plot and $\delta_{i}$ denotes its statistical
uncertainty. $N_i^{MC}(a_{K^{+}K^{-}})$ stands for the content of the same bin in generated plots. $\alpha$ is 
a normalization factor. $Re(a_{K^+K^-})$ and $Im(a_{K^+K^-})$ was varied in the range from 0 to 10 fm 
for data at an excess energy Q~=~10~MeV and in the range from 0 to 1 fm for the data at Q~=~28~MeV. \\
The method of least squares states that the best values of $a_{K^{+}K^{-}}$ are those, for which 
 $\chi^2$ is minimal.
\begin{figure}[H]
\centering 
\parbox{0.4\textwidth}{
{\epsfig{file=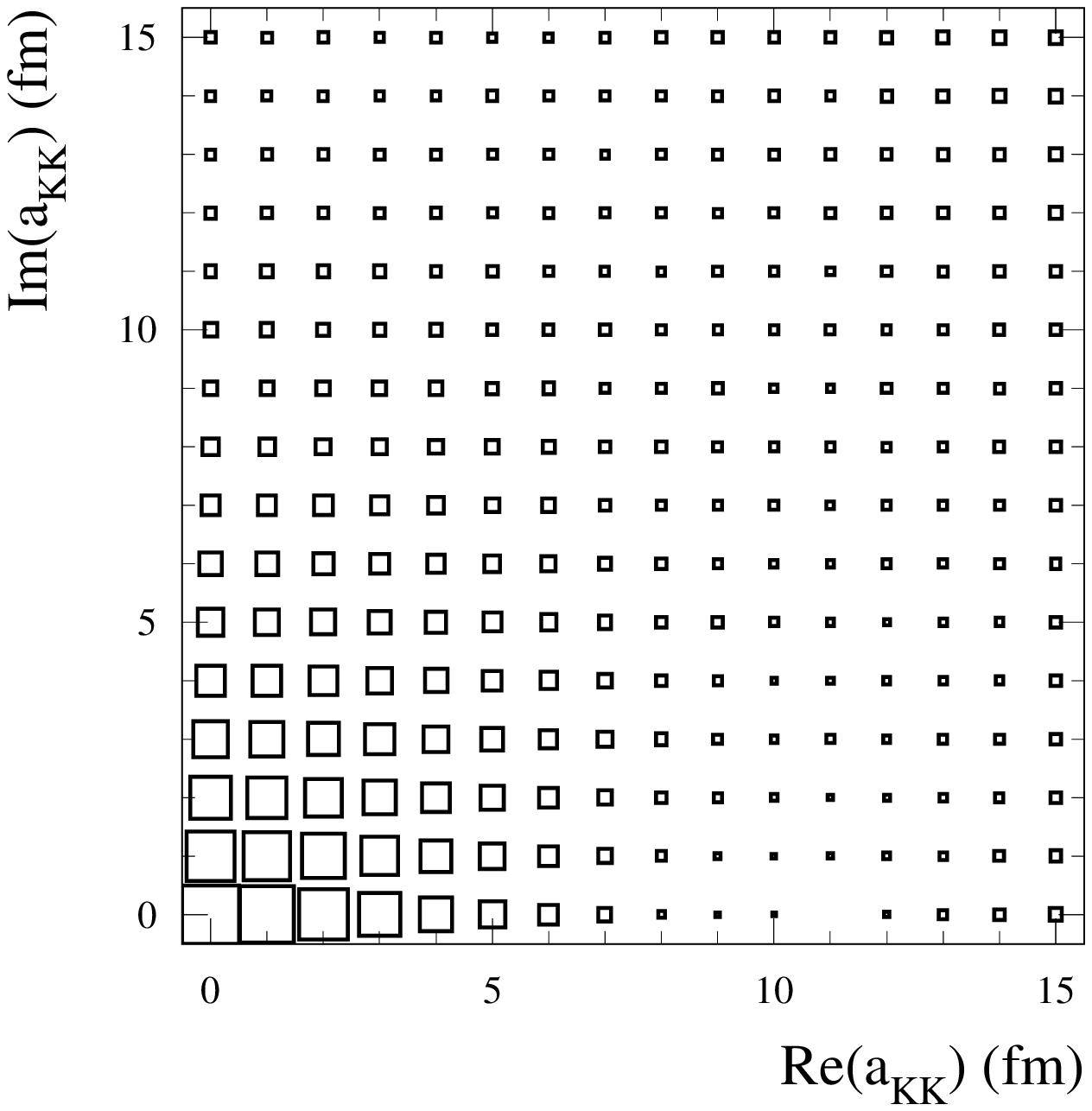,width=0.4\textwidth}}}
\parbox{0.4\textwidth}{
{\epsfig{file=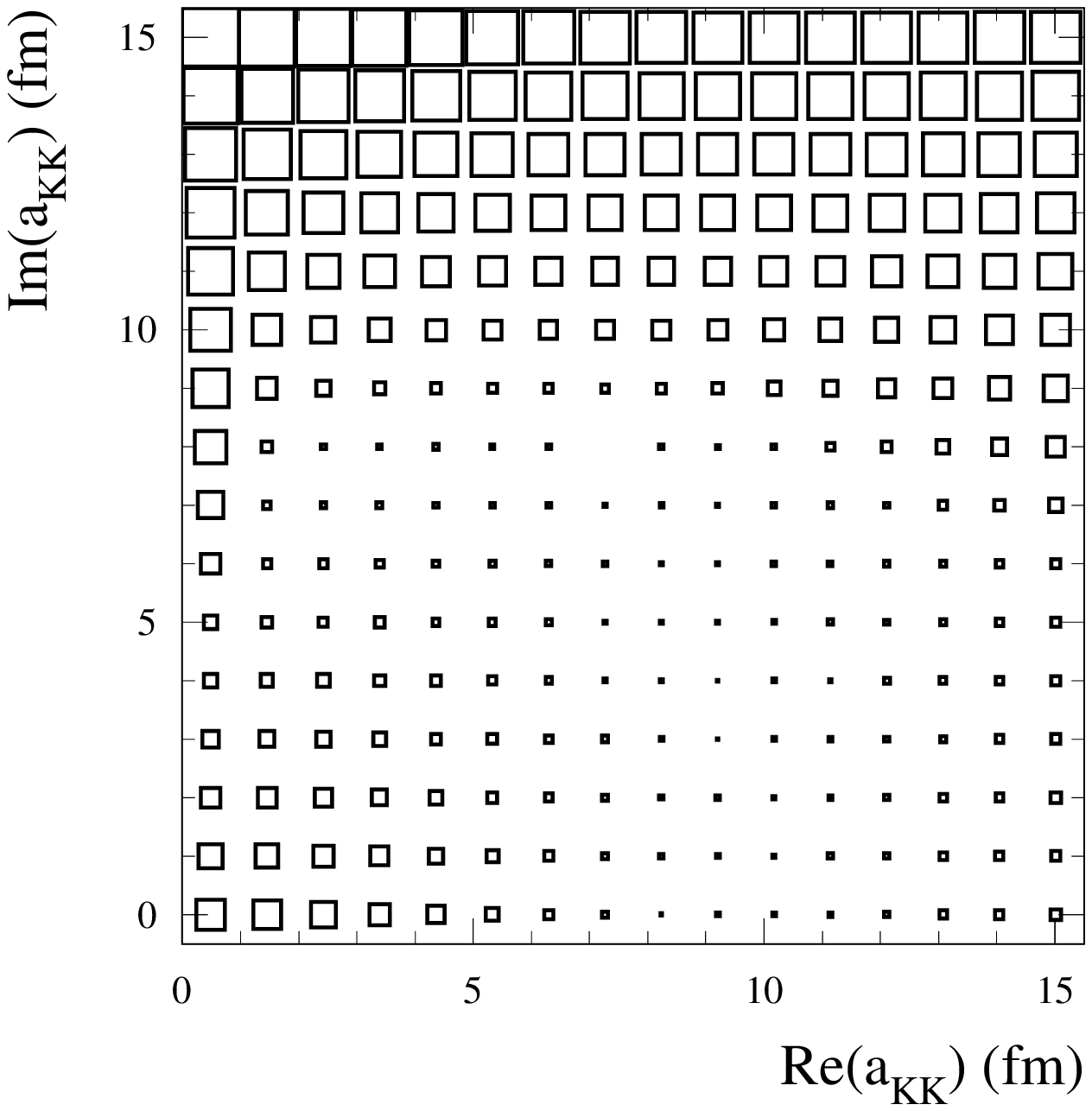,width=0.4\textwidth}}}
\parbox{0.4\textwidth}{
{\epsfig{file=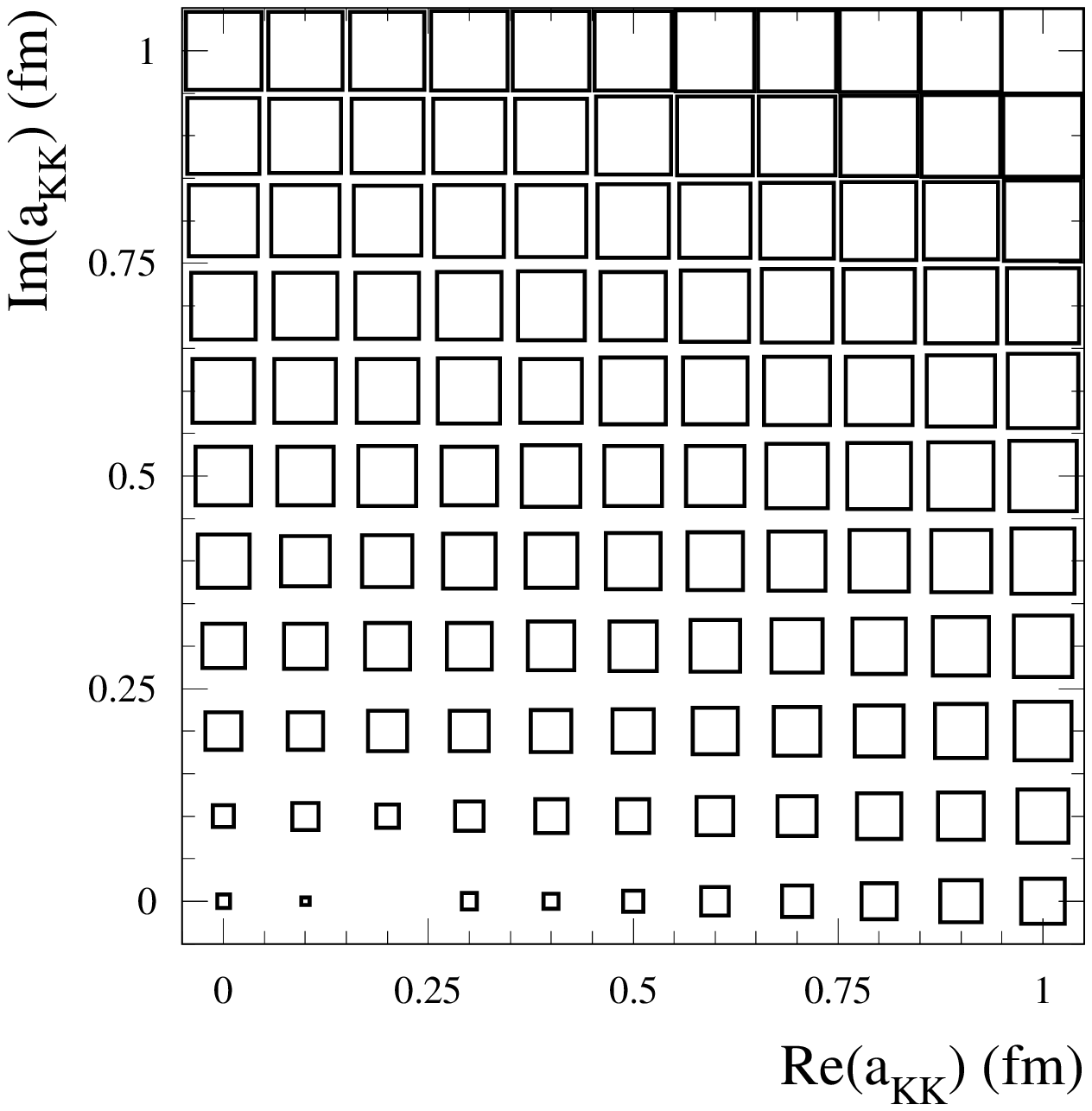,width=0.4\textwidth}}}
\parbox{0.4\textwidth}{
{\epsfig{file=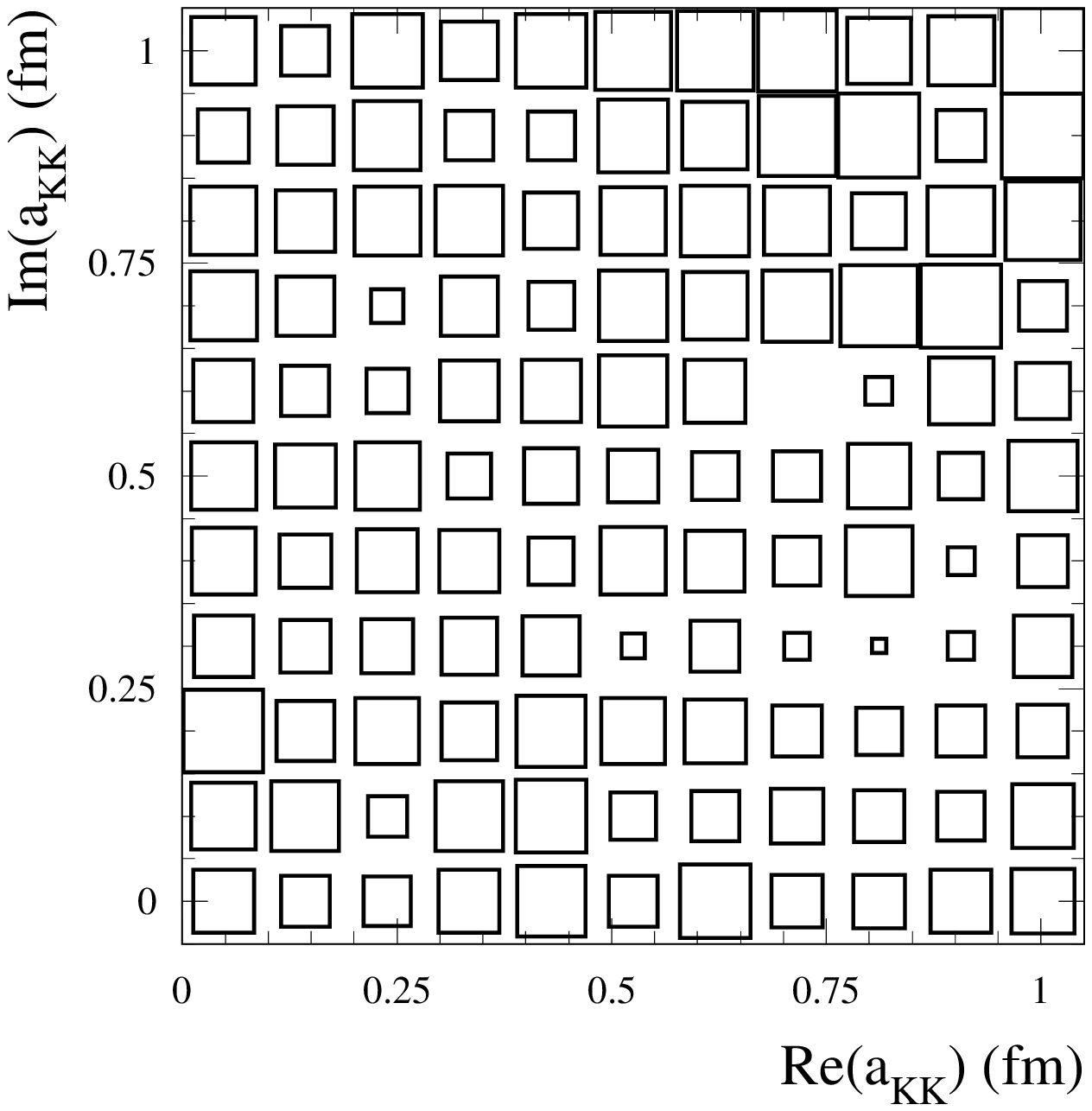,width=0.4\textwidth}}}
\caption{\textbf{(Upper panel)}$\chi^2$ as a function of $Re(a_{K^+K^-})$ and $Im(a_{K^+K^-})$ for Goldhaber (left plot) and Nyborg (right plot)
distributions at Q~=~10~MeV;~(\textbf{Lower panel)}~analogous $\chi^2$ distributions at Q = 28 MeV .
\label{chi2_1}
}
\end{figure} 
Fig.~\ref{chi2_1} presents distributions of the minimal $\chi^2$ calculated at both energies as a function of the real and imaginary part 
of the $K^+K^-$ scattering amplitude. At excess energy of Q~=~10 MeV for both Goldhaber and Nyborg plot 
the distribution has a minimum $\chi^2_{min}$ at $Re(a_{K^+K^-})$ = 11~$\pm$~8 fm and at $Im(a_{K^+K^-})$~= 0~$\pm$~6 fm . 
In case of Q~=~28~MeV we obtained:
\begin{center}
$a_{K^+K^-}$~=~(0.2~$\pm$~0.2)~+~$i$(0.0~$\pm$~0.5)~fm.
\end{center}
The statistical error was calculated by determination of the $a_{K^+K^-}$ interval within which the $\chi^2_{min}$ 
changes by $\Delta \chi^2$ = 1\cite{eadie}.
\pagestyle{myheadings}
\chapter{Summary and conclusions}
This thesis concerns the analysis of the $K^+K^-$ final state interaction which appears to be very important 
in many physical topics like for example the structure of the lightest scalar resonances. 
Measurements of the total cross section for the near threshold $pp\rightarrow ppK^{+}K^{-}$ reaction, 
performed by COSY-11 and ANKE collaborations, reveal a significant 
enhancement between the experimental data and theoretical expectations which neglect the 
interaction of the $K^+K^-$ pair. This observation encouraged us to extend the analysis of the 
$pp\rightarrow ppK^{+}K^{-}$ reaction into the differential cross sections. The investigations were based on 
experimental data determined by the COSY-11 collaboration at excess energies of Q~=~10~MeV and 28~MeV. 
For the purposes of the analysis we introduced two generalizations of the Dalitz plot for four particles 
in the final state proposed by Goldhaber and Nyborg. The experimental Goldhaber and Nyborg plots 
were compared to the results of Monte Carlo simulations generated with various values of the $K^{+}K^{-}$ 
scattering length. Beside the kaon-antikaon interaction in the simulations 
the interactions in $pp$ and $ppK^-$ subsystems were taken into account.\\ The values of the $K^+K^-$ scattering length, determined 
using the least squares method, amount to:
\begin{center}
$a_{K^+K^-}$~=~(11~$\pm$~8)~+~$i$(0~$\pm$~6)~fm for Q~=~10~MeV~,
\end{center}
and
\begin{center}
$a_{K^+K^-}$~=~(0.2~$\pm$~0.2)~+~$i$(0.0~$\pm$~0.5)~fm for Q~=~28~MeV~.
\end{center}
Unfortunately, due to low statistics, the standard deviations of the obtained values are rather big. The results 
indicate, that the $K^{+}K^{-}$ final state interaction is negligible which is especially seen for the analysis of the
data at Q~=~28~MeV. But also for 
the data closer to threshold the determined scattering length is consistent with zero within the standard 
statistical accuracy. This suggests that both the real and imaginary part of $a_{K^{+}K{-}}$ are 
rather small and amount to a fraction of fm.\\
It is worth mentioning, that there is another measurement of the $pp\rightarrow ppK^{+}K^{-}$ made by COSY-11 
at Q~=~4.5 MeV, which is at present under analysis\cite{gil}. Data obtained from this 
measurement may be very useful in the future to determine the kaon-antikaon scattering length more precisely.
\chapter{Acknowledgements}
I would like to express my gratitude to all the people who served me with their help during five years of 
my studies. I am especially grateful to:
\begin{itemize}
	\item Prof. dr hab. Paweł Moskal for plenty of time spent on introducing me to the particle kinematics and basis of the 
	Monte Carlo simulations, and for great patience in correcting my mistakes. He is the best teacher I have ever had.
	\item dr. Peter Winter for giving me access to the experimental data and valuable discussions about Monte Carlo simulations.
	\item Prof. Walter Oelert and dr. Dieter Grzonka for enabling me to visit the Research Centre J\"{u}lich, and for 
	reading and correcting this thesis.
	\item Prof. Colin Wilkin for explanation of the $pK^-$--FSI parameterization and valuable comments on this thesis.
	\item Prof. dr hab. Bogusław Kamys and Prof. dr hab. Lucjan Jarczyk for interesting lectures and useful comments during 
	ours seminars.
	\item all my colleagues from COSY-11 and WASA-at-COSY collaborations for their help and friendly atmosphere during 
	my stays in J\"{u}lich and meetings.
	\item mgr Dagmara Rozpędzik, Jarosław Zdebik, Marcin Zieliński and Jakub Bożek for the nice atmosphere of daily work.
	\item my parents the whole family for their great love 
	and trust placed in me. You were always by me when I needed your support.
	\item and finally to my sister Martyna and beloved Kasia Grzesik, for their incessant love and support.
\end{itemize}
\thispagestyle{empty}

\end{document}